\documentclass[twocolumn,preprintnumbers,amsmath,amssymb,a4paper]{revtex4}

\usepackage{graphicx}
\usepackage{bm}
\usepackage{color}
\usepackage{morefloats}
\usepackage{hyperref}
\usepackage{ulem}

\begin{document}


\title{The mean shape of transition and first-passage paths}

\author{Won Kyu Kim}
\email{wkkim@physik.fu-berlin.de}
\author{Roland R. Netz}%
        \email{rnetz@physik.fu-berlin.de}
\affiliation{Department of Physics, Freie Universit$\ddot{a}$t Berlin, Arnimallee 14, 14195 Berlin, Germany\\
}


\begin{abstract}
We  calculate the mean shape of  transition paths and first-passage paths based on the one-dimensional 
Fokker-Planck equation in an arbitrary free energy landscape including  a general inhomogeneous diffusivity profile. 
The transition path ensemble is the collection of all paths that do not revisit the start  position $x_A$ and that terminate
when first reaching the final position $x_B$. In contrast, a first-passage path can revisit but not cross its start position $x_A$ 
before it terminates at $x_B$.
Our   theoretical framework employs the forward and backward Fokker-Planck equations as well as
 first-passage, passage, last-passage and transition-path time distributions,
for which we derive the defining integral equations.
We show that  the mean time at which the transition path ensemble visits an intermediate position $x$ is equivalent to 
the mean first-passage time of reaching the starting position $x_A$ from $x$ without ever visiting $x_B$. 
The mean shape of first-passage paths is related  to the mean shape of transition paths by a constant time shift.
Since for large barrier height $U$  the mean first-passage time scales exponentially in  $U$ while the mean transition path time
scales linearly inversely in $U$, the time shift between first-passage and transition path shapes is substantial.
We present  explicit  examples of transition path shapes for linear and harmonic potentials
and illustrate our findings by trajectories generated from Brownian dynamics simulations.
 \end{abstract}

\maketitle

\section{Introduction}

For a reaction involving a free energetic  barrier,
the ensemble of transition paths is the collection of all paths that lead from the reactant to the  product ensemble without recrossing the boundaries
between the transition domain and the reactant domains \cite{Geissler_AnnuRevPhysChem02,Hummer_PNAS04,Metzner_JCP06}.
For continuous paths  described by the Fokker-Planck equation, transition paths can be generated by imposing absorbing boundary
conditions  on the boundaries between the reactant, transition and product domains \cite{Hummer_JCP_04}.
The mean transition path time $\tau^{TP}$
 is the first moment of the transition path time distribution. Based on an explicit formula derived by A. Szabo for the one-dimensional case \cite{Hummer_JCP_04},
$\tau^{TP}$  is for a large free-energetic barrier $U$ much shorter than Kramers' mean first-passage time $\tau^{KFP }$.
Note that a first-passage path is  allowed  to revisit its origin 
many times and in the Fokker-Planck description is obtained by imposing a reflecting boundary condition at the start position. In fact, while Kramers'
mean first-passage time grows exponentially with the energy barrier height $U$, the mean transition path time decreases linearly inversely in $U$
for a fixed separation between the start and final position along the 
one-dimensional  reaction coordinate \cite{Makarov10}.
This means that in a reaction involving a large energetic barrier, 
the system spends an exponential amount of time revisiting the reactant state, while the actual transition 
occurs very quickly \cite{rhoades2004two,Chung_PNAS09}.

Although transition paths are crucial for the understanding of rare events, 
they are in typical experiments that measure reaction rates not directly accessible. 
This situation dramatically changed with the advent of high resolution single molecule experiments that allow
to actually observe the folding and unfolding transition paths of 
proteins \cite{rhoades2004two,Chung_PNAS09,Chung24022012,Yu04092012,chung2013single}
as well as nucleic acid molecules \cite{lee2007measuring,PhysRevLett.109.068102,Truex2015}.
Note that in these experiments, reaction paths are typically obtained  from  the FRET efficiency 
between fluorophores connected to molecular positions that allow to 
separate folded from unfolded state. As such, these experiments project the complex molecular dynamics onto a one-dimensional reaction
coordinate that corresponds to an intramolecular distance, 
which motivated  extensive theoretical work using models restricted to  one-dimensional diffusion (though it is clear that a projection into one dimension
does not necessarily mean that a Markovian description is valid). 
 Indeed, in these experiments it was found that the mean transition path time
 is significantly smaller than the folding or unfolding time.
 In fact, the transition typically occurs so quickly that  only upper estimates can experimentally be obtained, which 
 from early estimates of about 
  $\tau^{TP} < 200 ~\mu$s for proteins as well as  RNA \cite{{rhoades2004two},{lee2007measuring},{Chung_PNAS09}}, 
   has come down to $\tau^{TP} < 10 ~\mu$s  with improved experimental time resolution  \cite{{Chung24022012}, {PhysRevLett.109.068102}, {Yu04092012}, {chung2013single}}.
  
The experimental advances created
 theoretical interest in transition paths and led to intense simulation activities \cite{  {shaw2010atomic},{Zhang18122012}, {frederickx2014anomalous}}
as well as the development of analytic approaches \cite{Orland07,Makarov10,Orland11}.
In this work, we present a theoretical framework for transition paths involving a combination of the backward Fokker-Planck equation,
the forward Fokker-Planck equation, and the renewal equation approach, and use  it to derive the mean shape of transition paths.
We use the same framework to also  calculate the mean shape of Kramers' first-passage paths. Interestingly, first-passage and transition
path shapes are identical modulo a shift by constant time which correspond to the residence time at the start position and is given by 
the difference of Kramers' mean first-passage time and the mean transition path time.
 We present  explicit results for transition path shapes for constant, linear and harmonic potentials and illustrate our findings with
 transition and first-passage paths generated using Brownian dynamics simulations.

\section{Derivation of Transition Path Times and  Shapes}
\label{sec_Intro}


The   Fokker-Planck (FP) operator  is defined as  \cite{weiss67,szabo_jcp80,zwanzig2001nonequilibrium}
\begin{equation}
 {\cal L}(x)ÔøΩ= \partial_x D(x)  {\rm e}^{- F (x)} \partial_x  {\rm e}^{ F (x)},
\end{equation}
where $F(x)$ is the free energy in units of the thermal energy $k_BT$ and $D(x)$
is the position-dependent diffusivity. In our previous analysis of protein folding trajectories from molecular dynamics trajectories
we   found that
the diffusivity profile has a pronounced spatial dependence, together with the free energy profile
it  allows to predict  kinetics that is rather insensitive 
on the precise definition of the reaction coordinate \cite{Hinczewski_JCP10}.
But even for the much simpler system of two water molecules diffusing relative to each other 
the diffusivity profile is not constant and therefore is important to take into account \cite{hansen2011}.
The Green's function can be formally written  as
\begin{equation}
{\cal G} (x,t | x_0) ={ \rm  e}^{t {\cal L}(x)}ÔøΩ\delta (x-x_0).
\end{equation}
It fulfills the initial condition
\begin{equation}
{\cal G} (x,0 | x_0) = ÔøΩ\delta (x-x_0),
\end{equation}
and solves the forward FP equation
\begin{equation} \label{FPE}
 \partial_t {\cal G} (x,t | x_0) =  {\cal L}(x) ÔøΩ{\cal G} (x,t | x_0).
\end{equation}
The  adjoint   FP operator \cite{weiss67,szabo_jcp80,zwanzig2001nonequilibrium}
\begin{equation}
 {\cal L}^\dagger (x_0)ÔøΩ=   {\rm e}^{ F (x_0)}   \partial_{x_0} D(x_0)  {\rm e}^{- F (x_0)} \partial_{x_0}   ,
\end{equation}
solves the backward FP equation
\begin{equation}  \label{BFPE}
 \partial_t {\cal G} (x,t | x_0) =  {\cal L}^\dagger(x_0) ÔøΩ{\cal G} (x,t | x_0).
\end{equation}
We will in the following section first use the backward FP equation, as it allows to derive transition path times
and first-passage times in a most transparent and direct fashion. We will then use the forward FP approach,
which requires careful normalization of expectation values but allows to calculate mean passage times and from that 
various relations between mean transition path, first-passage and passage times. Finally, we use the renewal equation approach
to derive constitutive relations between transition path time, first-passage time and last-passage time distributions. Here we will
be able to present a clear interpretation of the expression derived for the mean shape of transition and first-passage paths.

\subsection{Backward Fokker-Planck  approach}
\label{sec_back}

\subsubsection{First-passage  time distributions}

The derivation in this section uses concepts and techniques presented previously in \cite{szabo_jcp80,zwanzig2001nonequilibrium}.
By assuming absorbing boundary conditions at positions  $x_A$ and $x_B$ we  calculate first-passage times for paths  that start at $x_0$
with $x_A < x_0 < x_B$ and reach the boundaries for the first time. For this we define the survival probability
\begin{equation}
S(x_0,t) = \int_{x_A}^{x_B}  {\rm d} x \, {\cal G} (x,t | x_0),
\end{equation}
that the paths have not reached yet an absorbing boundary with the obvious properties $S(x_0,0)=1$ and, for regular free energies,
$S(x_0,\infty)=0$. The first-passage distribution for reaching either one of the boundaries is defined as
\begin{equation} \label{Kdef}
K(x_A \lor x_B, t | x_0) =- \partial_t S(x_0,t),
\end{equation}
and by using Eq. \eqref{FPE}  can be rewritten as
\begin{eqnarray}
K(x_A \lor x_B, t | x_0 ) &=& -   \int_{x_A}^{x_B}  {\rm d} x \,  {\cal L}(x) ÔøΩ{\cal G} (x,t | x_0) \nonumber\\
 &=&  \int_{x_A}^{x_B}  {\rm d} x \, \partial_x j (x,t | x_0) \nonumber\\
 &=& j (x_B,t | x_0) -   j (x_A,t | x_0),
\end{eqnarray}
where we used the flux at position $x$ defined as
\begin{equation} \label{flux}
 j (x,t | x_0) = - D(x)  {\rm e}^{- F (x)} \partial_x  {\rm e}^{ F (x)}   ÔøΩ{\cal G} (x,t | x_0) .
 \end{equation}
This shows that the total first-passage distribution can be decomposed into the two first-passage
distributions $K(x_A,t | x_0) = -  j (x_A,t | x_0)$ and $K(x_B,t | x_0) =   j (x_B,t | x_0)$ corresponding
to the respective boundary fluxes according to
\begin{equation}
K(x_A \lor x_B, t | x_0) = K(x_A,t | x_0)  + K(x_B,t | x_0) .
\end{equation}
By applying the flux operator defined in Eq. \eqref{flux} on both sides of the backward FP equation Eq. \eqref{BFPE} we obtain explicit equations for
the first-passage distributions $K(x_A,t | x_0)$  and $K( x_B,t | x_0 )$ as
\begin{equation} \label{passdist}
\partial_t K (x_{A/B},t | x_0) = {\cal L}^\dagger (x_0)ÔøΩ K(x_{A/B},t | x_0).
\end{equation}
Defining the n-th moments of the first-passage distributions as
\begin{equation}
K^{(n)} (x_{A/B}| x_0) =\int_0^\infty {\rm d}ÔøΩt \, t^n  K (x_{A/B},t | x_0) ,
\end{equation}
we obtain from Eq. \eqref{passdist} the set of equations
\begin{equation} \label{moments}
-n K^{(n-1)}(x_{A/B}| x_0) = {\cal L}^\dagger (x_0) K^{(n)} (x_{A/B}| x_0) ,
\end{equation}
where in the derivation we used the boundary condition that $ K (x_{A/B},t | x_0)=0 $
for $t=0$ and $t=\infty$. Thus all moments can be calculated recursively by straightforward
integration of Eq. \eqref{moments}.
The zeroth moment of the first-passage distribution is nothing but the
splitting probability,
\begin{equation} \label{eq:splitting0}
\phi_{A/B} (x_0) =  K^{(0)}(x_{A/B}| x_0),
\end{equation}
which gives the probability that  a path starting at $x_0$ reaches the boundary at $x_A$ or $x_B$.
From Eq. \eqref{moments} we obtain for $n=0$
\begin{equation} \label{splitting}
{\cal L}^\dagger (x_0) \phi_{A/B} (x_0) = 0.
\end{equation}
From Eq. \eqref{Kdef} and the boundary conditions $S(x_0,0)=1$ and $S(x_0,\infty)=0$
we conclude that $ \int_0^\infty {\rm d}ÔøΩt \,   [K(x_{A},t | x_0) + K(x_{B},t | x_0) ]=1$, in other words,
the sum of the splitting probabilities is unity, eventually the path reaches  a boundary,
\begin{equation}
 \phi_{A} (x_0)+  \phi_{B} (x_0) = 1.
\end{equation}
For $n=1$ we obtain  from Eq. \eqref{moments}
\begin{equation}  \label{Keq}
{\cal L}^\dagger (x_0) K^{(1)}(x_{A/B}| x_0)  = -  \phi_{A/B} (x_0).
\end{equation}
Since the first-passage distributions $K^{(1)}(x_{A}| x_0)$ and $K^{(1)}(x_{B}| x_0)$ are
 not normalized, reflected by the fact that  the splitting probabilities $\phi_{A/B} (x_0)$
are smaller than unity, the mean first-passage times are after normalization given by
\begin{equation} \label{tau1}
\tau^{FP} (x_{A/B}| x_0)  =
\frac{  K^{(1)} (x_{A/B}| x_0)   }{ \phi_{A/B} (x_0)}.
\end{equation}

As a side remark, the mean first-passage time to reach either the boundary $x_A$ or $x_B$ is given by the sum
of the first moments $ \tau^{FP} (x_A \lor x_B  | x_0) =K^{(1)} (x_{A}| x_0) + K^{(1)}(x_{B}| x_0) $. Adding the two equations for
$K^{(1)}(x_{A}| x_0)$ and $K^{(1)}(x_{B}| x_0)$ in Eq. \eqref{Keq} and using that $ \phi_{A} (x_0)+  \phi_{B} (x_0) = 1$ we arrive at
the familiar equation \cite{szabo_jcp80,zwanzig2001nonequilibrium}
\begin{equation}
{\cal L}^\dagger (x_0) \tau^{FP}  (x_A \lor x_B  | x_0)  = -1.
\end{equation}

\subsubsection{Splitting probabilities}\label{sec:splitp}

We explicitly show the calculation of the splitting probabilities, all further calculations proceed similarly and are not detailed.
We write  Eq. \eqref{splitting} explicitly for $\phi_B(x_0)$,
\begin{equation}
 {\rm e}^{ F (x_0)}   \partial_{x_0} D(x_0)  {\rm e}^{- F (x_0)} \partial_{x_0} \phi_B(x_0) = 0.
\end{equation}
Integrating once we obtain
\begin{equation}  \label{eq:diff_phiB}
 \frac{  {\rm e}^{ F (x)}   }{ D(x)}  =  C \partial_{x} \phi_B(x),
\end{equation}
where $C$ is an integration constant that will be determined later. Another integration yields
\begin{equation}   \label{phiB}
C^{-1}      \int_{x_A}^{x_0}ÔøΩ{\rm d}x \,   \frac{  {\rm e}^{ F (x)}   }{ D(x)}  =  \left.  \phi_B(x) \right|_{x_A}^{x_0} = \phi_B(x_0),
 \end{equation}
where we used that $ \phi_B(x_A)=0$, i.e., a path that starts at the absorbing boundary at $x_A$ will be immediately absorbed
and the probability to reach $x_B$ vanishes.  Conversely, $ \phi_B(x_B)=1$ and thus
\begin{equation}
C   =    \int_{x_A}^{x_B}ÔøΩ{\rm d}x \,   \frac{  {\rm e}^{ F (x)}   }{ D(x)} .
 \end{equation}
For  $ \phi_A(x_0)$ we obtain
\begin{equation}   \label{phiA}
 \phi_A(x_0) =1 -   \phi_B(x_0) = \frac{1}{C}    \int_{x_0}^{x_B}ÔøΩ{\rm d}x \,   \frac{  {\rm e}^{ F (x)}   }{ D(x)} .
  \end{equation}

\subsubsection{Mean first-passage times}

From Eq. \eqref{Keq} and using the results for $\phi_A(x_0)$ and $\phi_B(x_0)$ in Eqs.  \eqref{phiB} and \eqref{phiA}
we can straightforwardly calculate the first moments of the first-passage distributions. The boundary conditions require some thought:
The  mean first-passage time to reach either absorbing boundary,
$ \tau^{FP}  (x_A \lor x_B  | x_0) =K^{(1)}(x_{A}| x_0) + K^{(1)}(x_{B}| x_0)$, vanishes at the boundaries, i.e.,
$ \tau^{FP}  (x_A \lor x_B  |x_A)  =  \tau^{FP}  (x_A \lor x_B  |x_B)=0$. It  follows that both
 first moments $K^{(1)}(x_{A}| x_0) $ and $K^{(1)}(x_{B}| x_0)$
must individually vanish at the absorbing boundaries, i.e. $K^{(1)}(x_{A}| x_A) = K^{(1)}(x_{A}| x_B) =0$ and
$K^{(1)}(x_{B}| x_A) = K^{(1)}(x_{B}| x_B)=0$. With these boundary conditions we obtain
\begin{widetext}
\begin{equation}
 K^{(1)}(x_{A}| x_0) =  C   \phi_{B} (x_0)   \int_{x_0}^{x_B}ÔøΩ{\rm d}x \,  {\rm e}^{ - F (x)}  \phi_A^2 (x) +
  C   \phi_{A} (x_0)   \int_{x_A}^{x_0}ÔøΩ{\rm d}x \,  {\rm e}^{ - F (x)}  \phi_A (x)   \phi_B (x),
\end{equation}
and
\begin{equation}
 K^{(1)}(x_{B}| x_0)  =  C   \phi_{A} (x_0)   \int_{x_A}^{x_0}ÔøΩ{\rm d}x \,  {\rm e}^{ - F (x)}  \phi_B^2 (x) +
  C   \phi_{B} (x_0)   \int_{x_0}^{x_B}ÔøΩ{\rm d}x \,  {\rm e}^{ - F (x)}  \phi_A (x)   \phi_B (x).
\end{equation}

From Eq. \eqref{tau1} the mean first-passage time to reach boundary A when starting from $x_0$ reads
\begin{equation} \label{tauFPA}
 \tau^{FP}(x_{A}| x_0)  =  C \frac{  \phi_{B} (x_0)}{ÔøΩ \phi_{A} (x_0)}  \int_{x_0}^{x_B}ÔøΩ{\rm d}x \,  {\rm e}^{ - F (x)}  \phi_A^2 (x) +
  C      \int_{x_A}^{x_0}ÔøΩ{\rm d}x \,  {\rm e}^{ - F (x)}  \phi_A (x)   \phi_B (x),
\end{equation}
while the mean first-passage time to reach boundary B when starting from $x_0$ reads
\begin{equation} \label{tauFPB}
 \tau^{FP}(x_{B}| x_0)  =  C  \frac{  \phi_{A} (x_0)}{ÔøΩ \phi_{B} (x_0)}    \int_{x_A}^{x_0}ÔøΩ{\rm d}x \,  {\rm e}^{ - F (x)}  \phi_B^2 (x) +
  C     \int_{x_0}^{x_B}ÔøΩ{\rm d}x \,  {\rm e}^{ - F (x)}  \phi_A (x)   \phi_B (x).
\end{equation}
\end{widetext}
As we will show explicitly below,  because of reversibility,
the mean first-passage time $ \tau^{FP}(x_{A}| x_0) $ in fact equals  the mean-time a  transition path that starts at the  boundary
$x_A$ and ends at the boundary $x_B$ needs in order to reach the intermediate position $x_0$, it thus determines  the
mean shape of the transition path,
 \begin{equation}
 \tau_{shape}^{TP}(x_0 | x_A) = \tau^{FP}(x_{A}| x_0),\label{eq:shapeA}
 \end{equation}
 parameterized in terms of the mean time as a function of the position.
Likewise, the mean first-passage time $ \tau^{FP}(x_{B}| x_0)$ corresponds to the mean time a transition path that starts at boundary
$x_B$ and ends at boundary $x_A$ needs in order  to reach the intermediate position $x_0$,
\begin{equation}
 \tau_{shape}^{TP}(x_0 | x_B) = \tau^{FP}(x_{B}| x_0).\label{eq:shapeB}
 \end{equation}
 Note that the paths that contribute to the shape $\tau_{shape}^{TP}(x_0 | x_{A/B})$ revisit the position  $x_0$ multiple times,
 as will be illustrated later on when we present explicit Brownian dynamics paths.

\subsubsection{Transition path times}

The transition path time denotes the mean time a path takes to reach from the absorbing boundary $x_A$ to the other absorbing
boundary at $x_B$. It is thus defined by
\begin{equation}
\tau^{TP} (x_{B}| x_A)=  \tau^{FP}(x_B| x_0 \rightarrow x_A).
 \end{equation}
In the limit $x_0 \rightarrow x_A$ the first term in Eq. \eqref{tauFPB} vanishes and we obtain in agreement with Szabo's result \cite{{Hummer_JCP_04}}
\begin{equation}  \label{Szabo}
\tau^{TP} (x_{B}| x_A) =  C     \int_{x_A}^{x_B}ÔøΩ{\rm d}x \,  {\rm e}^{ - F (x)}  \phi_A (x)   \phi_B (x).
 \end{equation}
The same result is obtained from Eq. \eqref{tauFPA} by the limiting procedure
$\tau^{TP}(x_A| x_B) =  \tau^{FP}(x_A| x_0 \rightarrow x_B)$, reflecting that
transition paths are reversible, i.e. $\tau^{TP} (x_{B}| x_A) = \tau^{TP} (x_A| x_B)$.

\subsection{ Forward Fokker-Planck  approach}
\label{sec_for}


It is instructive to describe  transition paths also using  the forward FP equation \cite{{Hummer_JCP_04}}
as this allows to define passage and residence times and to
derive various useful relations between transition path times, first-passage times, and passage times. 

 Defining moments of the Green's function as
\begin{equation}
{\cal G}^{(n)} (x | x_0)  =  \int_0^\infty {\rm d}ÔøΩt \, t^n   {\cal G} (x,t | x_0),
\end{equation}
we obtain from the forward FP Eq. \eqref{FPE}  for $n>0$ the recursive relations
\begin{equation} \label{For1}
-n {\cal G}^{(n-1 )} (x | x_0)  =  {\cal L} (x)    {\cal G}^{(n)} (x | x_0).
\end{equation}
For $n=0$ we obtain
\begin{equation} \label{For2}
- \delta(x - x_0)  =  {\cal L} (x)    {\cal G}^{(0)} (x | x_0).
\end{equation}
We again impose absorbing boundary conditions at $x_A$ and $x_B$, i.e.
$ {\cal G} (x_A,t | x_0) =   {\cal G} (x_B,t | x_0) =0$, which means that all moments satisfy
$ {\cal G}^{(n)} (x_A | x_0)=  {\cal G}^{(n)} (x_B | x_0)=0$. 
Equations  \eqref{For1} and \eqref{For2}
are solved straightforwardly by integration, yielding
\begin{widetext}
\begin{equation} \label{For3}
 {\cal G}^{(0)} (x | x_0)  = C  {\rm e}^{ - F (x)}  \left\{   \phi_{A} (x_0)   \phi_B (x) - \theta(x-x_0) [  \phi_{A} (x_0)  -  \phi_{A} (x)]ÔøΩ\right\},
\end{equation}
and
\begin{equation} \label{For4}
 {\cal G}^{(1)} (x | x_0)  = C  {\rm e}^{ - F (x)}  \left\{   \phi_{A} (x)  \int_{x_A}^{x}ÔøΩ{\rm d}x' \,   {\cal G}^{(0)} (x' | x_0) \phi_B(x') +
                                                                                        \phi_{B} (x)  \int_{x}^{x_B}ÔøΩ{\rm d}x' \,   {\cal G}^{(0)} (x' | x_0) \phi_A(x')\right\},
 \end{equation}
\end{widetext}
where $ \theta(x-x_0)$ denotes the Heavyside function with the properties  $ \theta(x-x_0)=1$ for $x>x_0$ and zero otherwise.
Note that we assume the start and end positions $x_0$ and $x$ of the paths to be inside the absorbing boundary conditions, i.e.,
$x_A< x<x_B$ and $x_A< x_0<x_B$.
The mean time to reach the position $x$ when starting out from position $x_0$ follows from proper normalization as
\begin{equation} \label{For5}
      \tau^P (x | x_0) =    \frac  { {\cal G}^{(1)} (x | x_0)  }{   {\cal G}^{(0)} (x | x_0)},
  \end{equation}
we call this time the mean passage time and it is always larger than the mean first-passage time unless the target position is
an absorbing boundary.
        The mean passage time is the mean time to reach the target at position $x$, while allowing for multiple recrossing events.
We obtain for $x_0<x$ the result
\begin{eqnarray} \label{For6}
      \tau^P (x | x_0) &=&
      C  \frac{  \phi_{A} (x_0)}{ÔøΩ \phi_{B} (x_0)}    \int_{x_A}^{x_0}ÔøΩ{\rm d}x' \,  {\rm e}^{ - F (x')}  \phi_B^2 (x') \nonumber\\
&& +
  C     \int_{x_0}^{x}ÔøΩ{\rm d}x' \,  {\rm e}^{ - F (x')}  \phi_A (x')   \phi_B (x') \nonumber\\
&&  +
    C  \frac{  \phi_{B} (x)}{ÔøΩ \phi_{A} (x)}    \int_{x}^{x_B}ÔøΩ{\rm d}x' \,  {\rm e}^{ - F (x')}  \phi_A^2 (x'),\nonumber\\
   \end{eqnarray}
while for $x<x_0 $ we obtain
\begin{eqnarray} \label{For7}
      \tau^P (x | x_0) &=&
      C  \frac{  \phi_{A} (x)}{ÔøΩ \phi_{B} (x)}    \int_{x_A}^{x}ÔøΩ{\rm d}x' \,  {\rm e}^{ - F (x')}  \phi_B^2 (x') \nonumber\\
&& +
  C     \int_{x}^{x_0}ÔøΩ{\rm d}x' \,  {\rm e}^{ - F (x')}  \phi_A (x')   \phi_B (x') \nonumber\\
&& +
    C  \frac{  \phi_{B} (x_0)}{ÔøΩ \phi_{A} (x_0)}    \int_{x_0}^{x_B}ÔøΩ{\rm d}x' \,  {\rm e}^{ - F (x')}  \phi_A^2 (x').\nonumber\\
   \end{eqnarray}
Obviously, the two expressions are connected by the symmetry $  \tau^P (x | x_0) =    \tau^P (x_0 | x) $ that reflects the reversibility
of the underlying processes described by the FP equation. We note that this symmetry also holds when $x_0$ and/or
$x$ are located on the absorbing boundaries $x_A$ and $x_B$ . This symmetry also holds when we shift the absorbing boundary conditions to infinity,
i.e. for $x_A \rightarrow  - \infty $ and/or  $x_B \rightarrow \infty$, that is in the absence of absorbing boundary conditions.

The mean first-passage times in Eqs. \eqref{tauFPA} and \eqref{tauFPB}  follow from the passage times by the limiting procedures
\begin{equation} \label{tauFPA2}
 \tau^{FP}(x_A| x_0)  =    \tau^P (x \rightarrow x_A | x_0),
\end{equation}
and
\begin{equation} \label{tauFPB2}
 \tau^{FP}(x_B| x_0)  =    \tau^P (x \rightarrow x_B | x_0).
\end{equation}
The expression
\begin{eqnarray} \label{For8}
      \tau^P (x_0 | x_0) &=&
      C  \frac{  \phi_{A} (x_0)}{ÔøΩ \phi_{B} (x_0)}    \int_{x_A}^{x_0}ÔøΩ{\rm d}x' \,  {\rm e}^{ - F (x')}  \phi_B^2 (x') \nonumber\\
&& +
    C  \frac{  \phi_{B} (x_0)}{ÔøΩ \phi_{A} (x_0)}    \int_{x_0}^{x_B}ÔøΩ{\rm d}x' \,  {\rm e}^{ - F (x')}  \phi_A^2 (x'), \nonumber\\
   \end{eqnarray}
measures  the mean time a path stays at the starting position $x_0$,
we call this time the residence time.
By explicit consideration of the results in Eqs. \eqref{tauFPA}, \eqref{tauFPB}, \eqref{Szabo}, \eqref{For8}
it turns out that the transition path time $\tau^{TP}(x_B | x_A)$ in Eq. \eqref{Szabo}
is related to  the first-passage times of reaching the absorbing
boundaries at $x_A$ and $x_B$ from an intermediate position $x_0$ by subtracting the residence time,
\begin{equation} \label{tauTP2}
\tau^{TP}(x_B | x_A)  =  \tau^{FP}(x_{A} |ÔøΩx_0) +  \tau^{FP}(x_{B} |ÔøΩx_0)  -  \tau^P (x_0  | x_0).
\end{equation}
        This shows that a transition path time can be constructed by adding the mean first-passage times of two paths starting at an arbitrary position $x_0$ that reach the boundaries $x_A$ and $x_B$.
Since each path recrosses the starting position, the residence time $\tau^{P}(x_0 | x_0)$ has to be subtracted in order not to overcount these recrossing events.
By a tedious but straightforward calculation one can show that
\begin{equation} \label{tauTP3}
 \tau^{FP}(x_{B} |ÔøΩx_0)   -  \tau^P (x_0  | x_0)  =  \tau^{TP}(x_{B} | x_0)  =   \tau^{TP}(x_0 | x_B),
 \end{equation}
 holds for  the transition path time of going from $x_0$  to
$x_B$ or from $x_B$ to  $x_0$.
        Combining this with Eq. \eqref{tauTP2} we thus find
\begin{eqnarray}
\tau^{TP}(x_A | x_B) &=&  \tau^{TP}(x_A | x_0) + \tau^{TP}(x_B | x_0) + \tau^{P}(x_0 | x_0) \nonumber\\
        &=&  \tau^{FP}(x_{A} | x_0) +  \tau^{TP}(x_{0} | x_B).\label{tauTP4}
\end{eqnarray}
Equation \eqref{tauTP4} demonstrates that the transition path time from $x_A$ to $x_B$ can be decomposed
into the first-passage time starting from an intermediate position $x_0$ and the transition path time continuing to the other boundary.
Together with our definition for the shape of a transition path in Eq. \eqref{eq:shapeA}, we conclude
\begin{eqnarray}
\tau_{shape}^{TP}(x_0 | x_A)  &=&  \tau^{TP}(x_B | x_A) -  \tau^{TP}(x_B | x_0) \nonumber\\
        &=&  \tau^{TP}(x_0 | x_A) +  \tau^{P}(x_0 | x_0)   ,\label{eq:shape_00}
\end{eqnarray}
i.e., the mean shape of a transition path from $x_A$ to $x_0$ is the transition path time from $x_A$ to $x_B$ minus the transition path time from $x_0$ to $x_B$,
or, alternatively, the transition path from $x_A$ to $x_0$ plus the residence time at $x_0$.

Finally, and as mentioned before,
 the symmetry of mean passage times  $  \tau^P (x | x_0) =    \tau^P (x_0 | x) $ also holds when we move the point $x$ onto the absorbing boundary 
$x_A$, this turns the mean passage time $  \tau^P (x_A | x_0)$ into the mean first-passage time and we obtain $  \tau^{FP} (x_A | x_0) =    \tau^P (x_0 | x_A) $.
Combining this with the definition Eq. \eqref{eq:shapeA} we find 
\begin{equation}
\tau_{shape}^{TP}(x_0 | x_A)  =  \tau^{FP} (x_A | x_0) =  \tau^P (x_0 | x_A),\label{eq:shape_01}
\end{equation}
and we see that the  transition path shape corresponds to the mean passage time of paths that start from the absorbing boundary $x_A$.
Note that the formulas Eqs. \eqref{tauTP2}-\eqref{eq:shape_01} have been explicitly derived in the presence of absorbing boundaries at
positions $x_A$ and $x_B$, we will show in  the next section that similar relation can be derived from integral equations
for the distribution of passage times. 

\subsection{Renewal equation approach }
\subsubsection{First-passage time distribution}

Explicit expressions for the transition path time can also be derived within  the renewal equation approach
without referral to an explicit underlying diffusive model. The relations derived in this section are thus more general 
than the previous derivations which were based on the one-dimensional FP equation. 
Also, the present derivation allows to understand more deeply  in what sense the first-passage time $\tau^{FP}(x_A | x_0)$ 
in the presence of an absorbing boundary at $x_B$
can be interpreted as the shape of a transition path starting from $x_A$, $\tau_{shape}^{TP}(x_0 | x_A)$.
In this section we do not impose absorbing boundaries unless explicitly mentioned.
Although we use  a one-dimensional reaction coordinate, our results can be readily generalized to higher dimensions.

We start with the  renewal equation \cite{cox1962renewal,van1992stochastic} 
\begin{equation} \label{renew}
{\cal G} (x,t | x_0) = {\cal G}_{x'}  (x,t | x_0) +
 \int_0^t {\rm d} t' \,     {\cal G} (x,t-t' | x') K (x',t' | x_0),
\end{equation}
which can be viewed as a general definition of the first-passage time distribution  $K(x' ,t | x_0)$ and where
${\cal G}_{x'}  (x,t | x_0)$ denotes the Green's function in the presence of an absorbing boundary condition at $x'$.
It is an alternative more general definition than the one presented  in Eq. \eqref{passdist}.
The renewal equation 
states that the ensemble of all paths starting at time zero at $x_0$ and that are at position $x$ at time $t$ can be decomposed
into paths that never reach the absorbing boundary condition at $x'$ and paths that hit the boundary $x'$ for the first time at time $t'$
and from there on diffuse freely to $x$. By letting the position of the absorbing boundary $x'$ coincide with $x$ we obtain the special case
\begin{equation} \label{eq:conv1}
{\cal G} (x,t | x_0) = \int_0^t {\rm d} t' \,    {\cal G} (x,t-t' | x) K (x,t' | x_0) .
\end{equation}
In terms of the Laplace transform $ \tilde{\cal G} (x,\omega | x_0) = \int_0^\infty {\text d}t {\cal G} (x,t | x_0)   {\rm e}^{ - \omega t}  $
 Eq. \eqref{eq:conv1} becomes
\begin{equation}
\tilde{\cal G} (x,\omega | x_0) =   \tilde{\cal G} (x,\omega | x)  \tilde K (x,\omega | x_0) .
\end{equation}
Using  that moments can be  calculated from the Laplace transform by
\begin{equation}
{\cal G}^{(n)} (x| x_0) \equiv   \int_0^\infty dt \, t^n  {\cal G} (x,t | x_0)     =
(- \partial_\omega)^n   \tilde{\cal G} (x,\omega | x_0)|_{\omega=0},
\end{equation}
the normalized first moments are related by
\begin{eqnarray}
- \partial_\omega \ln \tilde K (x,\omega | x_0) \vert_{\omega = 0} &=& \frac{  K^{(1)} (x| x_0) }{ K^{(0)} (x| x_0)} \nonumber\\
 &=&
\frac{ {\cal G}^{(1)} (x| x_0) }{{\cal G}^{(0)} (x| x_0) }ÔøΩ-   \frac{ {\cal G}^{(1)} (x| x) }{{\cal G}^{(0)} (x| x) },\nonumber\\
\end{eqnarray}
or
\begin{equation} \label{decompose}
 \tau^{FP}(x|x_0)  =   \tau^P (x  | x_0)  -   \tau^P (x  | x) .
\end{equation}
In other words, the mean first-passage time $ \tau^{FP}(x|x_0)$ of going from $x_0$ to $x$
 in the absence of any additional absorbing or reflecting boundaries can be constructed from the
mean passage time $ \tau^P (x  | x_0)$ of going from $x_0$ to $x$ by subtracting the residence time $\tau^P (x  | x)$ of staying at $x$.
By symmetry of the passage time (derived in the previous section) we can write
\begin{equation}
 \tau^{FP}(x|x_0)  =   \tau^P (x_0  | x)  -   \tau^P (x  | x). \label{eq:tfpxx0}
\end{equation}
 This relation  holds also  in the presence of
 an absorbing boundary condition at $x_0$ (note that an absorbing boundary condition can 
 be simply imposed by creating a potential well of infinite depth in the region $x<x_0$, which turns
 $x_0$ into an absorbing boundary for all paths that come from $x>x_0$). 
 This turns $ \tau^{FP}(x|x_0) $ into the transition path time
 $ \tau^{TP}(x|x_0) $, the passage time $ \tau^P (x_0  | x)$ into the first-passage time
  $ \tau^{FP} (x_0  | x)$, and the residence time $\tau^P (x  | x)$ without specified boundary conditions
  into the residence time at $x$ in the presence of an absorbing boundary at $x_0$, which
  we denote by $ \tau^P_{x_0} (x  | x)$. We thus obtain from Eq. \eqref{eq:tfpxx0}
  \begin{equation} \label{TPgen1}
 \tau^{TP}(x|x_0)  =  \tau^{TP}(x_0 |x)  =   \tau^{FP} (x_0  | x)  -   \tau^P_{x_0} (x  | x),
\end{equation}
which is equivalent to  Eq. \eqref{tauTP3} (note that 
 Eq. \eqref{tauTP3}  by way of  derivation holds in the presence of two absorbing boundary conditions at $x_A$ and $x_B$, 
 so to make the equivalence perfect we can either shift the boundary $x_A$ in  Eq. \eqref{tauTP3} to infinity  or impose
 an additional absorbing boundary condition in Eq. \eqref{TPgen1}).

In order to derive Eq. \eqref{tauTP4} we need a convolution equation for first-passage times.
For this we choose in the renewal equation  \eqref{renew}  the absorbing  boundary condition $x'$ 
at a relative position $x_0 < x' < x$ and in this case obtain 
\begin{equation} \label{renew2}
{\cal G} (x,t | x_0) = 
 \int_0^t {\rm d} t' \,     {\cal G} (x,t-t' | x') K (x',t' | x_0).
\end{equation}
We now impose an absorbing boundary condition at $x$, which turns both Green's functions into first-passage time distributions 
so that we obtain
\begin{equation} \label{integralK}
 K (x,t | x_0) = \int_0^t {\rm d}  t' \,    K (x,t - t' | x') K (x',t' | x_0) ,
\end{equation}
valid for arbitrary positions $x'$ with $x_0 < x' < x$.
By using Laplace transformation, similarly as the calculation leading  to Eq. \eqref{decompose},  this yields
\begin{equation}\label{eq:gen_tau_fp}
 \tau^{FP}(x|x_0)  =   \tau^{FP} (x'  | x_0)  +  \tau^{FP} (x  | x') .
\end{equation}
Imposing an additional absorbing boundary condition at $x_0$ turns this into
\begin{equation}   \label{TPgen2}
 \tau^{TP}(x|x_0)  =   \tau^{TP} (x'  | x_0)  +  \tau^{FP}_{x_0} (x  | x'), 
\end{equation}
where the subindex $x_0$ in the last term indicates that an absorbing boundary is present at $x_0$.
This is identical  to Eq. \eqref{tauTP4}, remembering that Eq. \eqref{tauTP4} was derived
in the presence of an absorbing boundary at $x_B$.
We next combine Eqs. \eqref{TPgen1} and \eqref{TPgen2} and obtain
\begin{equation}   \label{TPgen3}
 \tau^{TP}(x|x_0)  =    \tau^{FP}_{x_0} (x  | x')  +  \tau^{FP} (x_0  | x')  -   \tau^P_{x_0} (x'  | x'),
\end{equation}
which is equivalent Eq. \eqref{tauTP2} if we impose an additional absorbing boundary condition at $x$.

\subsubsection{Transition path  time distribution}

We now impose an absorbing boundary condition at position $x_0$ in the convolution equation \eqref{integralK},
this turns the two first-passage time distributions starting at $x_0$ into transition path time distributions and we obtain 
\begin{equation} \label{integralT1}
 T (x,t | x_0) = \int_0^t {\rm d}  t' \,   K_{x_0} (x,t - t' | x')   T (x',t' | x_0), 
\end{equation}
where $K_{x_0} (x,t - t' | x')$ is the first-passage time distribution with an additional absorbing boundary condition at $x_0$ with
$x_0< x' < x$. Note that Eq. \eqref{TPgen2} follows directly from this integral equation via Laplace transformation.
It means that a transition path can be decomposed into a transition path to an intermediate position $x'$ followed by a
first-passage path from $x'$ that does not revisit $x_0$.

To go on with our derivation we define the last-passage distribution via the integral equation
\begin{equation} \label{lastpass1}
{\cal G} (x,t | x_0) = {\cal G}_{x'}  (x,t | x_0) +
 \int_0^t {\rm d} t' \,   H (x,t - t' | x')  {\cal G} (x',t' | x_0).
\end{equation}
In essence, the last-passage distribution $H (x,t' | x')$ comprises all paths that go from $x'$ to $x$ without revisiting the starting point at $x'$.
By moving the starting position $x_0$ to the absorbing boundary at $x'$ we obtain 
\begin{equation} \label{lastpass2}
{\cal G} (x,t | x') =
 \int_0^t {\rm d} t' \,   H (x,t- t' | x')  {\cal G} (x',t' | x').
\end{equation}
We now impose two absorbing boundary conditions, one at $x$ and the other at $x_0$ with the condition $x_0 < x' < x$, and obtain
\begin{equation} \label{lastpass3}
K_{x_0} (x,t | x') =
 \int_0^t {\rm d} t' \,   T (x,t - t' | x')  {\cal G}_{x_0,x} (x',t' | x').
\end{equation}
By inserting this integral equation into Eq. (\ref{integralT1}) we obtain
\begin{eqnarray} \label{integralT2}
 T (x,t | x_0) & = \int_0^t {\rm d}  t'  \int_0^{t-t'}  {\rm d}  t'' 
    \,  T (x,t - t' - t''  | x')  \nonumber �\\
 &  {\cal G}_{x_0,x} (x',t'' | x')    T (x',t' | x_0), 
\end{eqnarray}
which has a nice  intuitive interpretation:
a transition path from $x_0$ to $x$ can be decomposed into a transition path from $x_0$ to an arbitrary mid-point position
$x'$, a path that starts from $x'$ and returns to $x'$ without reaching the boundaries at $x_0$ and $x$, and finally
a transition path from $x'$ to the final destination $x$. 

We now use the renewal equation  (\ref{eq:conv1}) and   impose an absorbing boundary condition at $x_0$ and 
replace the variable $x$ by $x'$ to yield 
\begin{equation} \label{lastpass3}
H  (x',t | x_0) = \int_0^t {\rm d} t' \,   {\cal G}_{x_0} (x',t-t' | x')  T (x',t' | x_0),  
\end{equation}
which is an explicit  integral equation for the last-passage  time distribution. 
We now impose an additional absorbing boundary condition at $x$ with the ordering $x_0 < x' < x$ and obtain
\begin{equation} \label{lastpass4}
H_{x}  (x',t | x_0) = \int_0^t {\rm d} t' \,    {\cal G}_{x_0,x} (x',t-t' | x') T (x',t' | x_0)  .
\end{equation}
By comparison with Eq. (\ref{integralT2}) we obtain 
\begin{equation} \label{integralT3}
 T (x,t | x_0) = \int_0^t {\rm d}  t' \,   T (x,t - t' | x')     H_{x}   (x',t' | x_0) .
\end{equation}
Also this expression, from which we will derive the transition path shape, 
has an intuitive interpretation:
a transition path from $x_0$ to $x$ can be decomposed into a last-passage path from $x_0$ 
to an arbitrary mid-point position $x'$ followed by a transition path from $x'$ to the final destination $x$. 
Note that the last-passage paths   from $x_0$  to $x'$  do not visit the absorbing boundary condition $x$
which is indicated  by the subscript.

By construction, the integrand 
in Eq. (\ref{integralT3}) is the joint probability that  a transition path 
starting from $x_0$ and ending at $x$ has a duration of $t$  and  is at time $t'$ at the position $x'$.  This is so because
paths for  times later than $t'$ proceed on transition paths to $x$ and  
do not visit back to $x'$ and therefore do not contribute to the probability of being at $x'$.
The average shape of a transition path thus is obtained by averaging 
the$T (x,t - t' | x')     H_{x}   (x',t' | x_0)$  both over the  intermediate time $t'$ and transition path duration $t$.
We thus obtain for the shape of a transition path from $x_A$ to $x_B$
\begin{equation} \label{shape1}
\tau_{ shape}^{TP} (x | x_A) 
   =   \frac{      \int_0^\infty {\rm d}  t  \int_0^t {\rm d}  t' \,   t'  T (x_B,t - t' | x)     H_{x_B}   (x,t' | x_A)   }
 {      \int_0^\infty {\rm d}  t  \int_0^t {\rm d}  t' \,    T (x_B ,t - t' | x)     H_{x_B}   (x,t' | x_A)   }.
\end{equation}
By slightly rearranging we obtain 
\begin{equation} \label{shape2}
\tau_{ shape}^{TP} (x | x_A) 
   =   \frac{      \int_0^\infty {\rm d}  t' \,   t'      H_{x_B}   (x,t' | x_A)   }
 {      \int_0^\infty {\rm d}   t' \,      H_{x_B}   (x,t' | x_A)   } = \tau^P_{x_A, x_B} (x | x_A),
\end{equation}
and thus have derived the important result that the shape of a transition path is given by the passage time from an absorbing boundary
at $x_A$ to a midpoint $x$ in the presence of a second absorbing boundary at $x_B$, as presented  in Eq.  (\ref{eq:shape_01}).
We remind the reader of the relation  Eq.  (\ref{eq:shape_01}) which shows that because of the symmetry of passage times,
instead of averaging over paths that come from the absorbing boundary $x_A$, one can equally well average over 
first-passage paths that start from $x$ and that end at the boundary $x_A$, the latter ensemble is for simulations much more 
easy to implement and we will explicitly demonstrate  the equivalence of both ensembles  in our simulations.

\section{The shape of Kramers' first-passage paths}

Here we consider the mean shape of the Kramers' first-passage paths
defined as paths that start from a reflecting boundary and reach an absorbing boundary. 
We  basically repeat the derivation steps from  the previous section but replace the absorbing boundary 
condition at $x_A$ by a reflecting one. 
If we impose a reflecting boundary at position $x_A$ in the convolution relation for the first-passage distribution
Eq. (\ref{integralK}) we obtain
\begin{equation} \label{integralK2}
 K_{\tilde x_A}� (x_B ,t | x_A) = \int_0^t {\rm d}  t' \,    K_{\tilde x_A} (x_B ,t - t' | x) K_{\tilde x_A} (x,t' | x_A) ,
\end{equation}
where we denote a reflecting boundary condition by a subscript with a tilde and an adsorbing 
boundary condition by a subscript without a tilde. 

We next impose an absorbing boundary condition at $x_B$ and a reflecting boundary condition at $x_A$ in 
the integral relation for the last-passage distribution Eq. (\ref{lastpass2}) and obtain
\begin{equation} \label{lastpass4}
K_{\tilde x_A}  (x_B,t | x') =
 \int_0^t {\rm d} t' \,   T (x_B,t - t' | x)  {\cal G}_{\tilde x_A,x_B} (x,t' | x).
\end{equation}
By inserting this integral equation into Eq. (\ref{integralK2}) we obtain
\begin{eqnarray} \label{integralFP1}
 K_{\tilde x_A}� (x_B ,t | x_A)  & = \int_0^t {\rm d}  t'  \int_0^{t-t'}  {\rm d}  t'' 
    \,  T (x_B ,t - t' - t''  | x)  \nonumber �\\
 &  {\cal G}_{\tilde x_A,x_B} (x,t'' | x)   K_{\tilde x_A}�  (x,t' | x_A), 
\end{eqnarray}
which has a similar  interpretation as the corresponding result for an absorbing boundary condition at the origin in Eq. (\ref{integralT2}):
a Kramers' first-passage  path from $x_A$ to $x_B $ can be decomposed into a first-passage  path from $x_A$ to an arbitrary mid-point position
$x$, a path that starts from $x$ and returns to $x$ without reaching the absorbing boundary at $x_B$ and without 
crossing the reflecting boundary at $x_A$, and finally
a transition path from $x$ to the final destination $x_B$. 

We next impose an absorbing boundary condition  at $x_B$ and a reflecting boundary condition  at $x_0 = x_A$ on the definition of the first-passage distribution
Eq. (\ref{eq:conv1}), from which we obtain
\begin{equation} \label{renew4}
{\cal G}_{\tilde x_A,x_B}   (x,t | x_A) = \int_0^t {\rm d} t' \,    {\cal G}_{\tilde x_A,x_B}  (x,t-t' | x) K_{\tilde x_A} (x,t' | x_A).
\end{equation}
Comparison with Eq. (\ref{integralFP1}) gives the integral equation 
\begin{equation} \label{integralFP2}
 K_{\tilde x_A}� (x_B,t | x_A)  = \int_0^t {\rm d}  t'  
    \,  T (x_B,t - t'   | x)    {\cal G}_{\tilde x_A,x_B}   (x,t | x_A). 
\end{equation}
We now use similar arguments leading to our expression for the transition path shape in Eq. (\ref{shape2}):
The integrand in Eq. (\ref{integralFP2}) is the joint probability that a first-passage path 
starting from $x_A$ and ending at $x_B$ has a  duration of $t$ and  is at time $t'$ at position $x$.  
The average shape of a first-passage  path is obtained by averaging 
 over both intermediate time $t'$ and the first-passage path duration $t$,
we thus obtain for the mean shape of a Kramers' first-passage  path from $x_A$ to $x_B$
\begin{equation} \label{FPshape}
\tau_{ shape}^{KFP} (x | x_A) 
   =   \frac{      \int_0^\infty {\rm d}  t' \,   t'      {\cal G}_{\tilde x_A,x_B}    (x,t' | x_A)   }
 {      \int_0^\infty {\rm d}   t' \,       {\cal G}_{\tilde x_A,x_B}    (x,t' | x_A)   } = \tau^P_{\tilde x_A, x_B} (x | x_A).
\end{equation}
The only difference to the result for the transition path shape Eq. (\ref{shape2}) is that the absorbing
boundary condition   at $x_A$ is replaced by a reflecting boundary condition.

By Laplace transformation of Eq. (\ref{integralFP2}) we obtain (similarly  as when we derived
Eq. (\ref{decompose}) from Eq. (\ref{eq:conv1}))  
\begin{equation} \label{KFP4}
 \tau^{KFP}(x_B | x_A) =  \tau^{TP}(x_B | x) + \tau_{\tilde x_A, x_B}^{P}(x | x_A),  
\end{equation}
where we defined the Kramers' mean first-passage time as $ \tau^{KFP}(x_B | x_A)  = \tau_{shape}^{KFP}(x_B | x_A)$ 
and which is explicitly given by \cite{Gardiner}
\begin{equation}
\tau^{KFP}(x_B | x_A) = 
            \int_{x_A}^{x_B} {\rm d} x \frac{e^{F(x)}}{D(x)}\int_{x_A}^{x} {\rm d} x' e^{-F(x')}.\label{eq:kramerstime}
\end{equation}
By combining  Eq. \eqref{eq:shape_00}, Eq. \eqref{FPshape} and Eq. \eqref{KFP4} we  find
\begin{eqnarray}
\tau_{shape}^{KFP}(x | x_A) = \tau_{shape}^{TP}(x | x_A)   +  \tau^{P}_{\tilde x_A, x_B}(x_A | x_A), \label{eq:shapes}
\end{eqnarray}
showing that the mean shape of Kramers' first-passage paths
$\tau_{shape}^{KFP}(x | x_A)$ and the mean shape of transition paths
$\tau_{shape}^{TP}(x | x_A)$ are identical and shifted by a constant given by
$ \tau^{P}_{\tilde x_A, x_B}(x_A | x_A)$.
 This shift corresponds to the passage time at the reflecting boundary $x_A$
 and is according to Eq. \eqref{KFP4}�
 given by  $ \tau^{P}_{\tilde x_A, x_B}(x_A | x_A) = \tau^{KFP}(x_B | x_A) -  \tau^{TP}(x_B | x) $.

\section{Results for explicit potentials}



\begin{figure*}
\centering
\includegraphics[width = 0.85\textwidth]{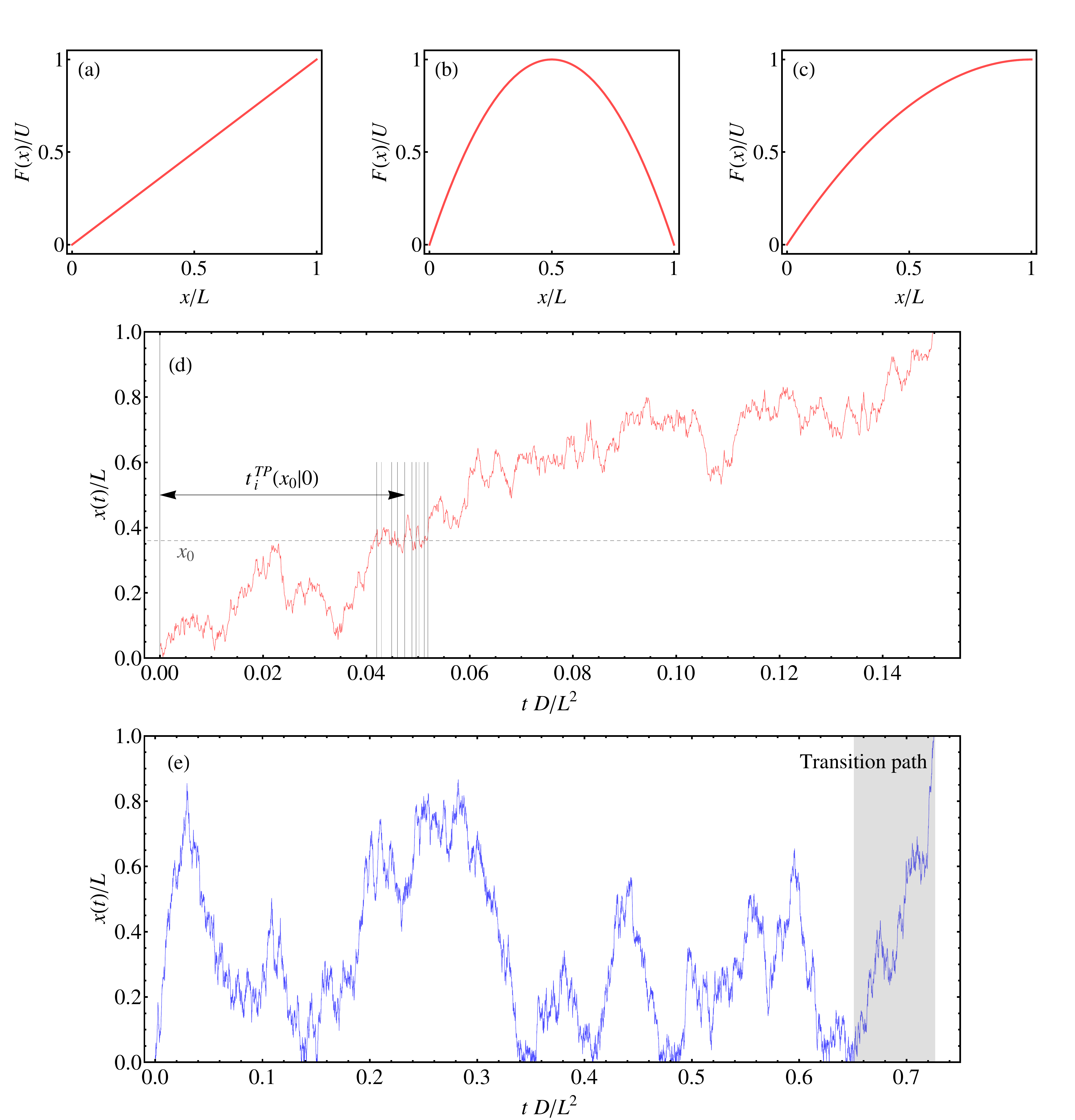}
\caption{Illustrations of the used rescaled  potentials ${F}({x})/U$ as a function of the rescaled length $x /L $, where $L$ is the transition length scale 
and $U$ is the barrier height: (a)  linear potential ${F}({x}) = U x/L$,
(b)  full harmonic potential ${F}({x}) = 4 U (1-x/L)x/L$
 and   (c)  harmonic ramp  ${F}({x}) = U (2-x/L)x/L$.
(d) A typical transition path trajectory $x^{TP}(t)$ for the  force-free case, obtained from Brownian dynamics (BD) simulations.
The times $t^{TP}_{i}({x}_0 | 0)$ when the transition path crosses the position $x_0$   are indicated by vertical lines.
(e) A typical Kramers' first-passage path  trajectory $x^{KFP}(t)$ for the force-free case, obtained from BD simulations, in the presence
of a reflecting boundary condition at $x=0$ and an absorbing boundary condition at $x=L$. The transition path is the last part of the trajectory indicated by the gray region.
}\label{fig:illust}
\end{figure*}

We next present exemplaric  transition path shapes  for a few different simple potential shapes shown in Fig. \ref{fig:illust}-(a)-(c).
We consider a reaction coordinate $x$ in the range of $0 \leq x \leq L$, where $L$ is the transition length scale,
and restrict ourselves from now on to a homogeneous diffusion constant $D$

\subsection{Brownian dynamics simulations and trajectory analysis}

We also present trajectories obtained from one dimensional overdamped Brownian dynamics (BD) simulations.
The simulations are based on the Langevin equation
\begin{eqnarray}\label{eq:bdeq0}
{\text{d}x(t) \over \text{d}t} = - D {\text{d}  {F}(x) \over \text{d} {x}} + {\zeta(t) \over \gamma},
\end{eqnarray}
where $\gamma = k_B T / D$ is the friction constant and $\zeta(t)$ is a Gaussian random force which fulfills 
$\langle \zeta(t) \rangle =0$ and $\langle \zeta(t) \zeta(t') \rangle = 2 \gamma k_B T  \delta(t-t')$.
	The  discretized and  rescaled Langevin equation reads
	\begin{eqnarray}\label{eq:bdeq}
\tilde{x}(\tilde{t}+\text{d}\tilde{t})= \tilde{x}(\tilde{t} ) - {\text{d} {F} \over \text{d}\tilde{x}} \text{d}\tilde{t} + \sqrt{2 \text{d}\tilde{t}}~r(\tilde{t}),
\end{eqnarray}
where $\tilde{x}=x/L$ is the rescaled position,  $\tilde{t} = t D/L^2$ is the rescaled time, and $r(\tilde{t})$ is a Gaussian random number 
 with zero mean and unit standard deviation.
We iterate Eq. \eqref{eq:bdeq}  with a typical time step $\text{d}\tilde{t} = 10^{-4}$.

To obtain mean first-passage times ${\tau}^{FP}(0 | {x}_0 )$ and  ${\tau}^{FP}(L | {x}_0 )$  we vary the initial position from ${x}_0 = 0$ to ${x}_0 = L$ 
and measure the time needed to
reach one of the two absorbing boundaries ${x}_A=0$ or ${x}_B=L$ for the first time, 
we typically average over $10^5$ first-passage times.

We also generate transition path trajectories within  BD simulations. In practice we  initiate a trajectory at a reflecting boundary at $x=0$ and record until it 
reaches the  absorbing boundary at $x=L$, the transition path trajectory is the last  portion of the trajectory after it has last returned to the reflecting boundary at $x=0$,
as shown in Fig. \ref{fig:illust}-(e). The mean transition path shape is obtained by averaging the time transition paths take to reach a certain position $x_0$
\begin{eqnarray}\label{eq:aveall}
{\tau}^{TP}_{shape}({x}_0|0)= \sum_{i=1}^{N} \frac{t^{TP}_{i}({x}_0|0)}{N},
\end{eqnarray}
where $t^{TP}_{i}({x}_0|0)$ denotes the time at which a transition path  trajectory that starts out at $x=0$ crosses the position $x_0$, as illustrated in 
Fig. \ref{fig:illust}-(d). Note that a single transition path crosses the position $x_0$ multiple times, the averaging in Eq. \eqref{eq:aveall} is done over 
the entire transition path ensemble and over all crossing events, $N$ thus counts the total number of crossing events in the entire transition path ensemble.
For our final results we typically generate $10^4$ transition paths.

In a similar manner, we analyze Kramers' first-passage trajectories, which start from a reflecting boundary at $x=0$ and eventually reach the  absorbing boundary at $x=L$,
an example of which is shown in  Fig. \ref{fig:illust}-(e). 
To obtain the mean shape of Kramers' first-passage trajectories, denoted by $\tau_{shape}^{KFP} (x_0 | 0)$,
we average  the mean time it takes such a path to reach a certain position $x_0$
	\begin{equation}
\tau^{KFP}_{shape}({x}_0|0) =  \sum_{i=1}^{N} \frac{t^{KFP}_{i}({x}_0|0)}{N},\label{eq:aveall_kfp}
\end{equation}
where $t^{KFP}_{i}({x}_0|0)$ denotes the time at which a path that starts from $x=0$ crosses $x=x_0$.

\subsection{Force-free case}

We first consider the force-free case, ${F}=0$. The splitting probabilities read ${\phi}_A(x) = 1 - x/L$ and ${\phi}_B(x) = x/L$
with ${C}=1$, and the transition path time according to Eq. \eqref{Szabo} reads
\begin{eqnarray}
{\tau}^{TP}(L|0)
&=&{L^2 \over 6D},
\label{eq:tp_zero1}
\end{eqnarray}
which is three times smaller than Kramers' mean first-passage time
\begin{eqnarray}
{\tau}^{KFP}(L|0) = {L^2 \over 2D},
\end{eqnarray}
        according to Eq. \eqref{eq:kramerstime}.
This decrease is due to the  subtraction of the part of the Kramers' first-passage trajectories that 
contains multiple returns to the origin, as illustrated in  Fig. \ref{fig:illust}-(e).

\begin{figure}
\centering
\includegraphics[width = 0.5\textwidth]{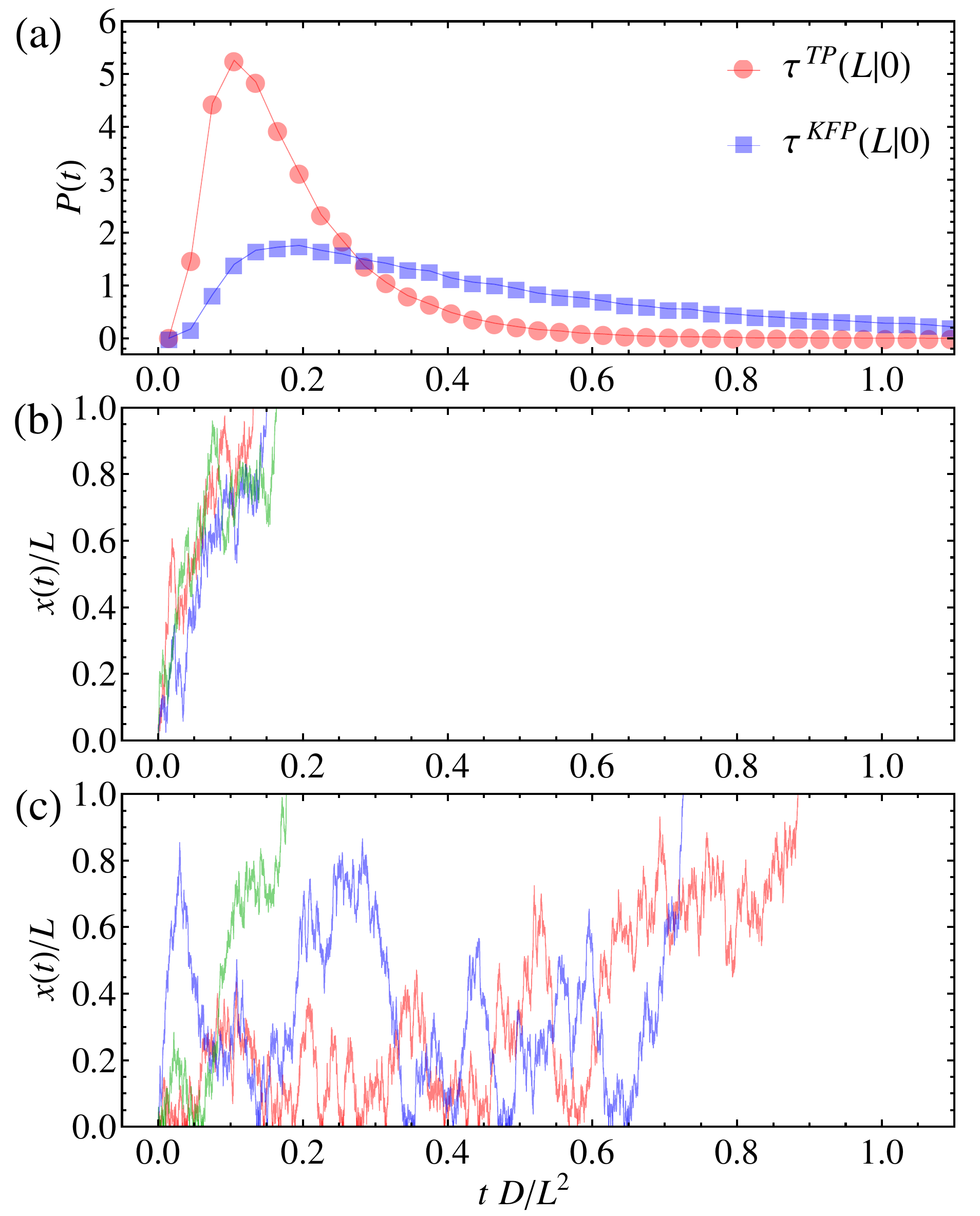}
\caption{
(a) The normalized distribution functions for the transition path time $\tau^{TP}(L|0)$ (circles) and the Kramers' first-passage time $\tau^{KFP}(L|0)$ (squares) in the force-free case, obtained from BD simulations.
(b) Three typical transition path trajectories $x(t)$.
(c) Three typical Kramers' first-passage trajectories  $x(t)$.
}\label{fig:zero_pdf}
\end{figure}

The normalized distribution functions for the transition path time $\tau^{TP}(L|0)$ (circles) and the Kramers' first-passage time $\tau^{KFP}(L|0)$ (squares) are shown in Fig. \ref{fig:zero_pdf}-(a), obtained from BD simulations.
The transition path time distribution is more  sharply peaked compared with the Kramers' first-passage time distribution. 
        The trajectories  shown in Fig. \ref{fig:zero_pdf}-(b) and (c) reflect this difference  of the two distributions.

\begin{figure}
\centering
\includegraphics[width = 0.45\textwidth]{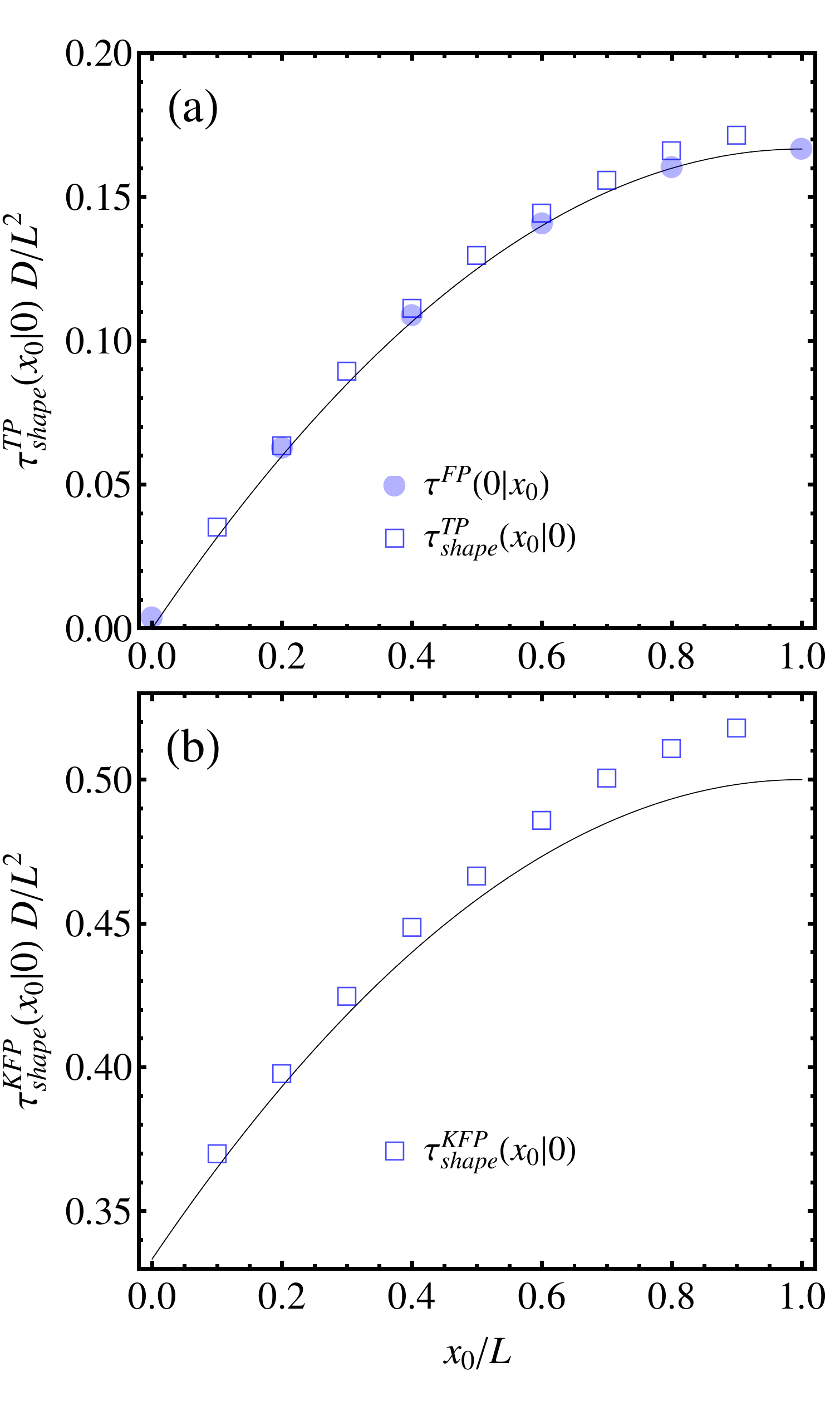}
\caption{
(a) Mean shape of transition paths  ${\tau}_{shape}^{TP}({x}_0 | 0)$  in the force-free case.
	The solid line shows the analytic result Eq. \eqref{eq:zero1}.
Filled circles show BD simulation results for the mean first-passage time
${\tau}^{FP}(0 | x_0)$ while open squares show the mean shape from the analysis of transition paths according to Eq. \eqref{eq:aveall}.
(b) Mean shape of Kramers' first-passage paths ${\tau}_{shape}^{KFP}({x}_0 | 0)$  in the force-free case.
Symbols show BD simulation results  while the solid line shows analytic results according to Eqs. \eqref{eq:shapes} and \eqref{eq:zero1}.
Note that the two curves in (a) and (b) are identical except a vertical shift by a constant time.}
\label{fig:zero_shape}
\end{figure}

The mean transition path shape is, according to Eqs. \eqref{tauFPA} and  \eqref{eq:shapeA}, given as
\begin{equation}
{\tau}_{shape}^{TP}({x}_0 |0) = {\tau}^{FP}(0| {x}_0) = \frac{L {x}_0}{6D}  \left( 2- \frac{{x}_0}{L} \right),\label{eq:zero1}
\end{equation}
and is depicted in Fig. \ref{fig:zero_shape}-(a) by  a solid line. Note that the transition path shape is a quadratic function, transition paths start out with finite
velocity  at the origin and reach the final destination with infinite velocity.
This asymmetry, which is a universal property of mean transition path shapes for all potentials, can be easily understood by considering
Eq. \eqref{eq:shape_00} and realizing that a mean  transition path time scales quadratic with the diffusion length scale in the limit of small diffusion length scale.
The filled symbols in Fig. \ref{fig:zero_shape}-(a) show the BD simulation results for the
first-passage time  ${\tau}^{FP}(0| x_0) $
while the open square symbols  show the BD results for  ${\tau}^{TP}_{shape}({x}_0|0)$ obtained via  Eq. \eqref{eq:aveall},
both simulation results  agree well with 
 the theoretical result Eq. \eqref{eq:zero1}.

The solid curve in Fig. \ref{fig:zero_shape}-(b) shows the Kramers' mean first-passage shape $\tau^{KFP}_{shape}(x_0 |0)$, 
calculated from Eqs. \eqref{eq:shapes} and \eqref{eq:zero1}.
The Kramers' mean first-passage shape $\tau^{KFP}_{shape}(x_0 |0)$ is, according to Eqs. \eqref{KFP4} and \eqref{eq:shapes}, 
identical to the transition path shape $\tau^{TP}(x_0|0)$ shifted by the amount $\tau^{P}_{\tilde{x}_A=0,x_B=L}(x) = \tau^{KFP}(L|0) - \tau^{TP}(L|0) = L^2/(3D)$. 
The  symbols in Fig. \ref{fig:zero_shape}-(b) show the BD results using Eq. \eqref{eq:aveall_kfp},
again, the agreement is very good. 


\subsection{Transition path in linear potential}


For a linear potential ${F}= U  {x}/L$ we find  for the transition path time
\begin{eqnarray}
{\tau}^{TP}(L|0)
&=&\frac{L^2}{D} \frac{U \coth \left(\frac{U}{2}\right)-2}{U^2},
\label{eq:tp_lin}
\end{eqnarray}
which is an even function of $U$. This means that the transition path time is the same irrespective of whether the transition paths go up the linear
potential or whether they go down. This of course follows directly from the  general symmetry of passage times in 
Eqs. \eqref{For6} 
and \eqref{For7} 
but is worthwhile pointing out again at this point.  
To leading order in $U$ the asymptotic behavior reads
\begin{eqnarray}
{\tau}^{TP} D/L^2
&\approx&
\begin{cases}
{1 \over 6}-{U^2 \over 360}~&,~|U| \ll 1\\
{1 / |U|}~&,~|U| \gg 1.
\end{cases}
\label{eq:tp_lin_assym}
\end{eqnarray}
\begin{figure*}
\centering
\includegraphics[width=\textwidth]{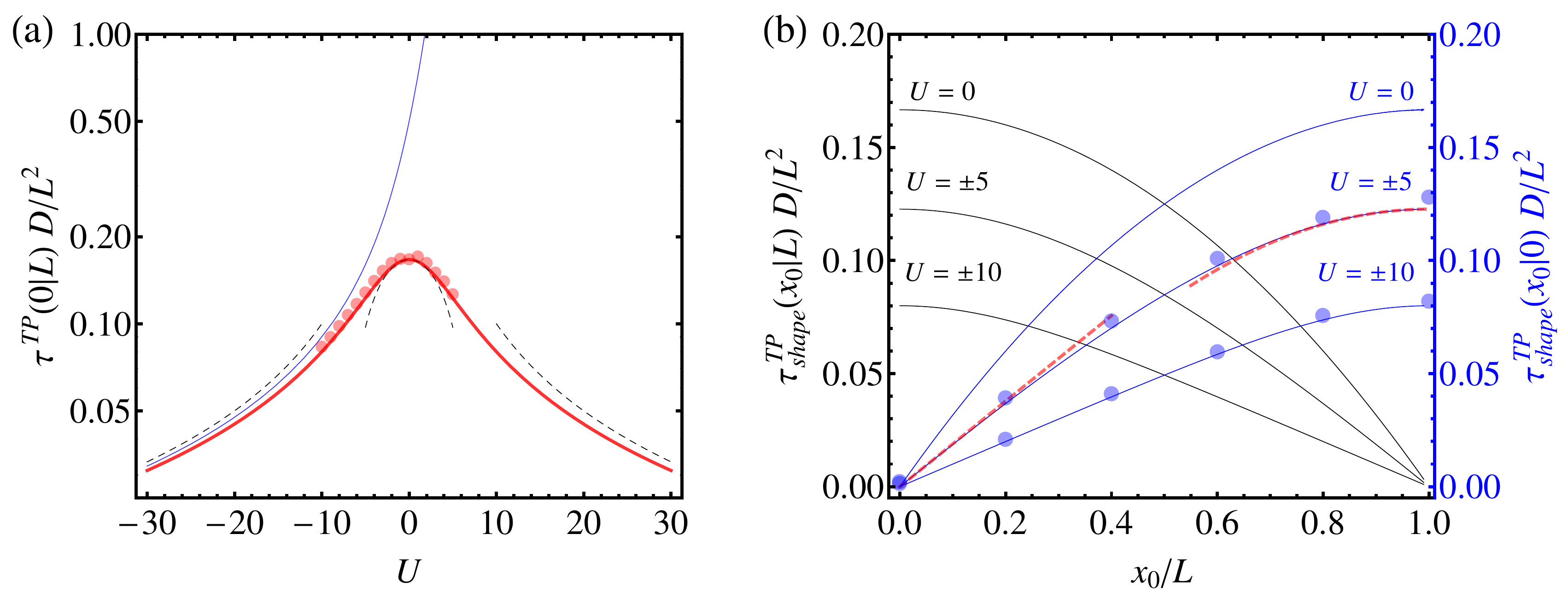}
\caption{\label{fig:4}
Results for a linear potential ${F}= U  {x}/L$.
(a) The solid red curve shows the mean transition path time ${\tau}^{TP}(0|L)$  from Eq. \eqref{eq:tp_lin}  as a function of $U$ on a log-linear scale.
 The broken curves depict the asymptotic expressions from Eq. \eqref{eq:tp_lin_assym}. For comparison, 
 the solid blue curve shows  Kramers' mean first-passage time ${\tau}^{KFP}(L|0) $ which monotonically increases with $U$.
 The symbols denote BD simulation results.
(b) Mean shapes of transition paths ${\tau}_{shape}^{TP}$.
Blue curves depict ${\tau}_{shape}^{TP}({x}_0|0)$ starting from the left boundary from Eq. \eqref{eq:lin1}, while black curves depict 
${\tau}_{shape}^{TP}({x}_0 |L)$ starting from the right boundary from Eq. \eqref{eq:lin2}.
Symbols denote  BD simulation results for $U=-5$ and $U=-10$ while
broken red curves depict the asymptotic  expressions for $U=\pm 5$ from Eq. \eqref{eq:lin_as}.
}
\end{figure*}
The red solid curve in Fig. \ref{fig:4}-(a) shows ${\tau}^{TP}(L|0)$ in Eq. \eqref{eq:tp_lin} while  the asymptotic  expressions in Eq. \eqref{eq:tp_lin_assym} 
are depicted by broken curves.
Note that the Kramers' mean first-passage time ${\tau}^{KFP}(L|0)=L^2(e^U -1 -U)/(DU^2)$, shown by a solid blue curve in Fig. \ref{fig:4}-(a),
shows very different behavior and in particular is a monotonically increasing function of $U$. For large potential strength $U\gg1$ we find 
 an exponential increase to leading order, ${\tau}^{KFP}(L|0) \sim e^U/U^2$.
The symbols in Fig. \ref{fig:4}-(a) show  BD simulation results for the transition path time, which agree  well with the theory.

A further noteworthy fact is that the mean  transition path time ${\tau}^{TP}$ is
for non-zero values of $U$ strictly  smaller than the force-free result ${\tau}^{TP}= L^2/(6D)$ corresponding to the maximum value obtained for  $U=0$.
This means that transition paths in a linear potential are faster  than  force-free transition paths,  regardless of 
whether the slope is positive or negative. 

\begin{widetext}


The transition path shapes read

\begin{eqnarray}
{\tau}_{shape}^{TP}({x}_0|0) 
&=& \frac{L^2}{D} \frac{\text{csch}\left(\frac{U}{2}\right) \text{csch}\left(\frac{U}{2}-\frac{U {x}_0}{2 L}\right)  \left[({x}_0 /L-2) \sinh \left(\frac{U {x}_0}{2 L}\right)+\frac{{x}_0}{L} \sinh \left(U-\frac{U {x}_0}{2 L}\right)\right]}{2 U},\label{eq:lin1}\\
{\tau}_{shape}^{TP}({x}_0|L) &=& \frac{L^2}{D} \frac{\coth(U/2) - \frac{{x}_0}{L} \coth(\frac{U {x}_0}{2L})}{U},\label{eq:lin2}
\end{eqnarray}
where ${\tau}_{shape}^{TP}({x}_0|0) $ has the asymptotic limits
\begin{eqnarray}
{\tau}_{shape}^{TP}({x}_0|0) D/L^2\approx
\begin{cases}
\frac{U - \sinh U}{U(1-\cosh U)}\frac{{x}_0}{L} &, ~{x}_0 \ll L\\
{\tau}^{TP}(L|0) D/L^2 -\frac{1}{6}(\frac{{x}_0}{L} - 1)^2 &, ~{x}_0 \approx L.
\end{cases}
\label{eq:lin_as}
\end{eqnarray}
\end{widetext}
 Figure \ref{fig:4}-(b) shows the transition path shapes ${\tau}_{shape}^{TP}$ as function of the position ${x}_0$, where the blue curves depict ${\tau}_{shape}^{TP}({x}_0|0) $ in Eq. \eqref{eq:lin1}, and the black curves depict ${\tau}_{shape}^{TP}({x}_0|L)$ in Eq. \eqref{eq:lin2}.
Symbols denote  BD simulation results for $U=-5$ and $U=-10$.
 The broken red curves depict the asymptotic  limits in Eq. \eqref{eq:lin_as} for $U=\pm 5$.
 Due to the symmetry of passage times, the shapes ${\tau}_{shape}^{TP}$ are symmetric   with respect to an exchange of starting  positions.

\subsection{Harmonic potential} \label{sec:tpt_harmonic}


For a harmonic potential ${F}= 4 U {x}(1-{x}/L) /L$ we find
for the transition path time

\begin{eqnarray}
{\tau}^{TP}(L|0) 
&=&
{ L^2 \over 4D} F_{2,2}(-U) \nonumber\\
&& - {L^2 \over 2 D \sqrt{\pi} U \text{erf}(\sqrt{U})} \int_{0}^{\sqrt{U}} \text{d} y ~ y^2 e^{-y^2} F_{2,2}(-y^2),\nonumber\\
\label{eq:tp_harm3}
\end{eqnarray}
where $F_{2,2}(x) = F_{2,2}(\{1,1\};\{3/2,2\};x)$ is the generalized hypergeometric function.
For the small barrier limit $| U | \ll 1$ we find to leading order
\begin{eqnarray}
{\tau}^{TP}(L|0)
\approx \frac{L^2}{D} \left[ 
{1 \over 6}-{2 \over 45}U \right],
\label{eq:tp_harm_assym1}
\end{eqnarray}
 which decreases from the force-free transition path time ${\tau}^{TP}=L^2/(6D)$.
For the  large barrier limit $U \rightarrow \infty$ we recover  the known asymptotic result \cite{Chung_PNAS09,Makarov10}
\begin{eqnarray}
{\tau}^{TP}(L|0)
\approx
{L^2 \ln(2 e^\gamma U) \over 8 D U},
\label{eq:tp_harm_assym2}
\end{eqnarray}
where $\gamma \approx 0.577$ is the Euler gamma constant, and we used 
$\text{erf}(\sqrt{U}) \approx 1$, $F_{2,2}(-U) \approx \ln(4 e^\gamma U)/(2U)$ and 
$\int_{0}^{\infty} \text{d} y ~ y^2 e^{-y^2} F_{2,2}(-y^2) = (\sqrt{\pi}/4)\ln(2)$ for large $U$.
 We note that the denominator $8U$ in Eq. \eqref{eq:tp_harm_assym2} can be reinterpreted as 
 the rescaled curvature $\omega^2 = L^2 \left| \left( \text{d}^2 {F}/ \text{d} {x}^2 \right)_{{x}=L/2} \right|$ at the barrier top of the harmonic potential,
 yielding the previously published form \cite{Chung_PNAS09}
\begin{eqnarray}\label{eq:rsc_szabo}
{\tau}^{TP}(L|0)  \approx  { L^2 \ln(2 e^\gamma U) \over  D \omega^2}.
\end{eqnarray}
For fixed potential curvature and varying potential height,
 Eq. \eqref{eq:rsc_szabo} shows that the transition path time increases logarithmically with increasing potential height $U$,
while for fixed diffusion $L$,
 Eq. \eqref{eq:tp_harm_assym2} shows that the transition path time decreases inversely linearly with increasing potential height $U$\cite{Makarov10}.

\begin{figure*}
\centering
\includegraphics[width=\textwidth]{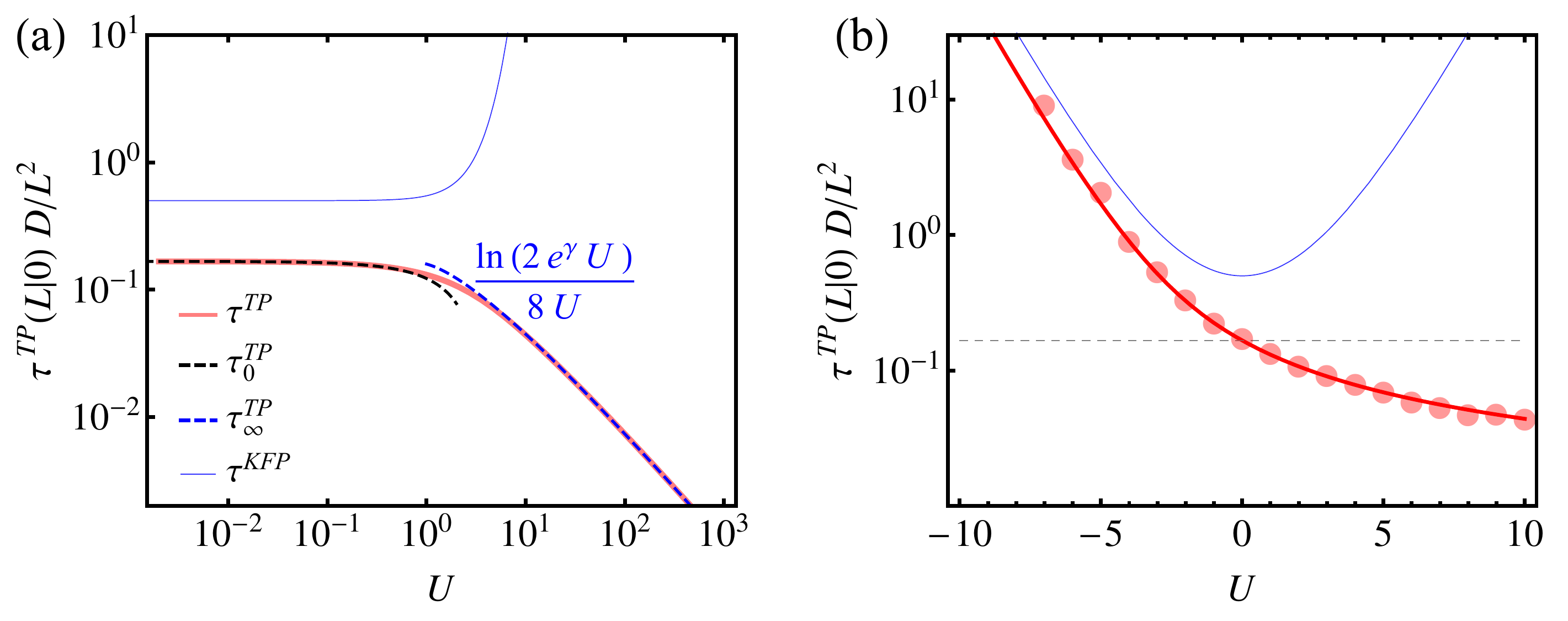}
\caption{\label{fig:ham_tpt}
Results  for  the harmonic potential ${F} = 4 U {x}(1-{x}/L)/L$ as a function of the barrier height $U$.
(a) Mean transition path time ${\tau}^{TP}(L|0)$ from  Eq. \eqref{eq:tp_harm3} (solid red curve) on a log-log scale, compared
with the asymptotic expressions Eqs. \eqref{eq:tp_harm_assym1} and  \eqref{eq:tp_harm_assym2} (dashed lines).
(b) Mean transition path time ${\tau}^{TP}(L|0)$  (solid red curve) on a log-linear compared with  BD simulation data (symbols).
Solid blue curves depict the Kramers' mean first-passage time ${\tau}^{KFP}(L | 0) $ from
 Eq. \eqref{eq:kfp_harmonic0}.       
 The horizontal dashed line depicts the force-free transition path time ${\tau}^{TP} = L^2/(6D)$.
}
\end{figure*}

In  Fig. \ref{fig:ham_tpt} we present ${\tau}^{TP}(L|0)$ as a function of the barrier height $U$.
In Fig. \ref{fig:ham_tpt}-(a) we show ${\tau}^{TP}(L|0) D/L^2$ from Eq. \eqref{eq:tp_harm3} on a log-log scale (solid red curve), 
which is seen to decrease from the force-free case ${\tau}^{TP}D/L^2=1/6$ as $U$ increases.
We also show  the asymptotic  expressions  Eqs. \eqref{eq:tp_harm_assym1} and  \eqref{eq:tp_harm_assym2} by dashed curves.
In Fig. \ref{fig:ham_tpt}-(b) we show ${\tau}^{TP}(L | 0)$ from Eq. \eqref{eq:tp_harm3} on a log-linear scale (solid red curve),
here we also compare with BD simulation results obtained via Eq. \eqref{eq:aveall}.
 The solid blue curves in Fig. \ref{fig:ham_tpt} depict the Kramers' mean first-passage time given by
\begin{eqnarray}
{\tau}^{KFP}(L | 0)
&=&
\frac{L^2}{D} \frac{\pi  \text{erf}\left(\sqrt{U}\right) \text{erfi}\left(\sqrt{U}\right)}{8 U}, \label{eq:kfp_harmonic0}
\end{eqnarray}
where $\text{erf}\left( x \right) = \frac{2}{\sqrt\pi}\int_0^x e^{-t^2}\,\mathrm dt$ is the error function, and $\text{erfi}\left( x \right) = \frac{2}{\sqrt\pi}\int_0^x e^{t^2}\,\mathrm dt$ is the imaginary error function.
The leading order result for large $|U|$ reads
\begin{eqnarray} \label{eq:kfp_harm}
{\tau}^{KFP}(L | 0) D/L^2&=&
 \frac{\sqrt{\pi}}{8}\frac{e^{|U|}}{|U|^{3/2}}=
 \frac{\sqrt{\pi}}{\omega^2}\frac{e^{|U|}}{\sqrt{|U|}} \label{eq:kfp_harmonic_1}.
\end{eqnarray}

In Fig. \ref{fig:ham_tpt}  we see that
the transition path time ${\tau}^{TP}(L|0)$ is a monotonically decreasing function of the barrier height $U$, 
while the Kramers' time ${\tau}^{KFP}(L | 0)$ is a symmetric function and has a minimum of ${\tau}^{KFP} = L^2/(2D)$  at $U = 0$. 
In fact, transition paths over a harmonic barrier with $U>0$ are faster, while transition paths over a harmonic well characterized
by $U<0$ are slower compared to the force-free case with $U=0$. 
This can be rationalized by Eq. \eqref{tauTP4},
since the transition path time for reaching from the boundaries to the center of the harmonic potential 
are rather insensitive on whether $U$ is positive or negative (as will be shown in the next section), but the residence time
at the center of the harmonic potential  is much larger for the case of a harmonic well with $U<0$ than 
for a harmonic barrier with $U>0$.  
The symmetric behavior of the Kramers' mean first-passage time can be understood based on Eq. \eqref{eq:gen_tau_fp}
since first-passage time are transitive: the first-passage time for traversing a harmonic potential is the sum of the 
first-passage times from the boundary to the middle and from the middle to the other boundary. 
We reiterate that mean first-passage times are transitive, as shown in  Eq. \eqref{eq:gen_tau_fp},
while transition path times are not, as shown in Eq. \eqref{tauTP4}.

\begin{figure}
\centering
\includegraphics[width=0.5\textwidth]{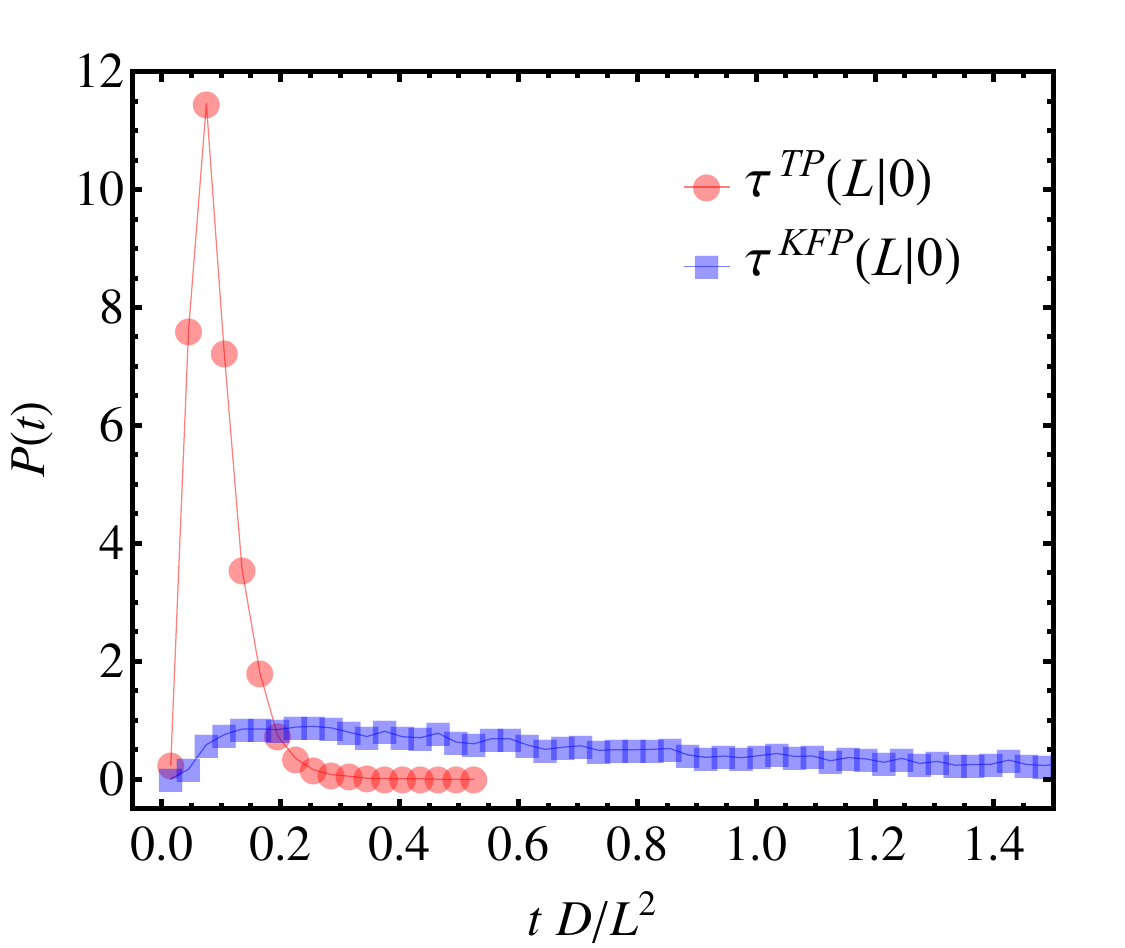}
\caption{\label{fig:pdf_harmonic}
Normalized distribution functions for the transition path time $\tau^{TP}(L|0)$ (circles) and the Kramers' first-passage time $\tau^{KFP}(L|0)$ (squares) in a
 harmonic potential at $U=3$, obtained from BD simulations.
}
\end{figure}

In Fig. \ref{fig:pdf_harmonic}
we show the normalized distribution functions for the transition path time  (circles) and for the Kramers' first-passage time (squares) for $U=3$, 
obtained  from BD simulations.
The transition path time  distribution shows a pronounced  peak around $\tau D/L^2 = 0.1$, 
 close to the mean transition path time $\tau^{TP}(L|0)(U=3) D/L^2 \approx 0.1$, as seen in Fig. \ref{fig:ham_tpt}.
 In contrast, the Kramers' first-passage time distribution   is quite broad, the first moment is given by  $\tau^{KFP}(L | 0) (U=3) D/L^2\approx 1$ and thus
  is 10 times  larger than the mean transition path time.

\begin{widetext}




For  the transition path shape we find
\begin{eqnarray}
{\tau}_{shape}^{TP}({x}_0 | 0)
&=&
{\tau}^{TP}(L | 0)-{L^2 \over 2 D U}\int_{\sqrt{U}}^{\sqrt{U}(2{x}_0/L-1)}\text{d}y\left( {\text{erf}(y) - \text{erf}(\sqrt{U}) \over \text{erf}(\sqrt{U}(2{x}_0/L-1)) - \text{erf}(\sqrt{U})} - {1 \over 2}\right)D_{+}(y)
, \label{eq:harm_shape2}
\end{eqnarray}
where $D_{+}(x)=e^{-x^2}\int_{0}^{x}\text{d}t e^{t^2}$ is the Dawson integral function. The second term in Eq. \eqref{eq:harm_shape2} vanishes for
 ${x}_0 = L$ and reduces to $-{\tau}^{TP}(L | 0)$ given in Eq. \eqref{eq:tp_harm3} for  ${x}_0 = 0$.
\end{widetext}

\begin{figure}
\centering
\includegraphics[width=0.5\textwidth]{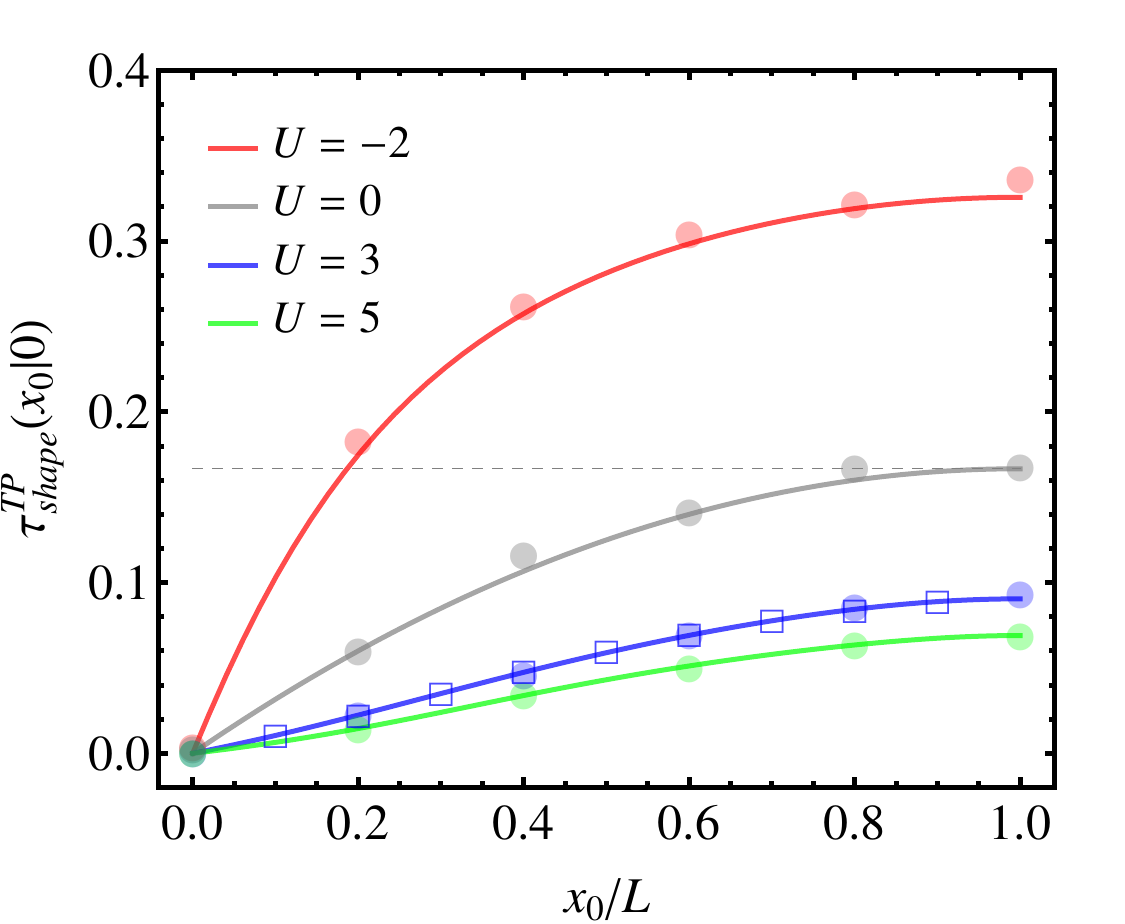}
\caption{\label{fig:harm_tpt_shape} Mean transition path shape ${\tau}_{shape}^{TP}({x}_0 | 0)$ from  Eq. \eqref{eq:harm_shape2},
for different values of the barrier height $U$ of the harmonic potential ${F} = 4 U {x}(1-{x}/L)/L$.
Symbols show BD simulation results for ${\tau}^{FP}(0 | x_0)$  (filled circles)
  and    ${\tau}_{shape}^{TP}({x}_0 | 0)$ (open squares) for $U=3$.      
   The horizontal dashed line depicts the force-free transition path time ${\tau}^{TP} = L^2/(6D)$.
}
\end{figure}

 Figure \ref{fig:harm_tpt_shape} depicts the mean transition path shapes ${\tau}_{shape}^{TP}({x}_0 | 0)$ in Eq. \eqref{eq:harm_shape2}
 for different values of the barrier height $U$. Transition paths are faster for positive values of $U$, i.e. for paths that have to go over a harmonic barrier top,
 while the slow down for negative values of $U$, i.e. for paths that have to traverse a harmonic well. Again, we observe a pronounced asymmetry  of the
 mean shape of transition paths, paths start out quickly and reach the boundary at $x=L$ with vanishing slope.
  Filled symbols show BD simulation results for ${\tau}^{FP}(0 | x_0)$ 
  while open symbols show  BD simulation results for  ${\tau}_{shape}^{TP}({x}_0 | 0)$, both for $U=3$. We observe good agreement between
  the two different ways of extracting transition path shapes, as expected based on our analytical results, as well as with our analytically derived
  shape.

\subsection{Harmonic ramp} \label{sec:tpt_half_harmonic}

Here we consider the harmonic potential ${F}({x}) = U {x}(2-{x/L})/L$ which has a barrier top $F=U$ at the final position 
$x=L$.

\begin{figure*}
\centering
\includegraphics[width=\textwidth]{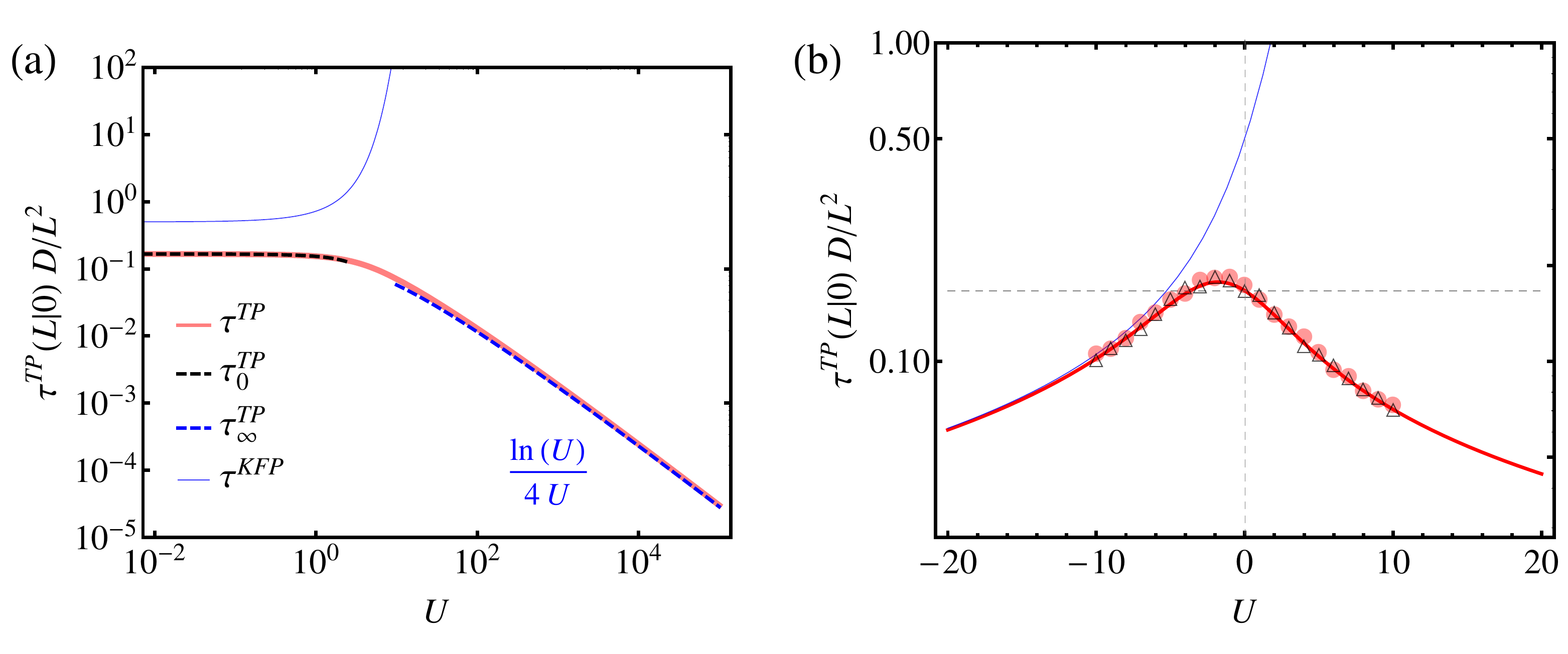}
\caption{\label{fig:ham_tpt_half}
Results for the harmonic ramp ${F}({x}) = U {x}(2-{x/L})/L$.
(a) Mean transition path time  ${\tau}^{TP}(L |0)$ from  Eq. \eqref{eq:tp2_half_harm} (solid red curve) on log-log scales,
the asymptotic  expressions  Eqs. \eqref{eq:tp_half_harm_ap1} and  \eqref{eq:tp_half_harm_ap2}  are shown by dashed black lines.
 The blue line shows Kramers' mean first-passage time ${\tau}^{KFP}(L | 0) D/L^2$ from Eq. \eqref{eq:kfp_half_harm}.
(b)   Same curves shown on log-linear scales,  compared with BD simulation data 
 for transition paths starting from the left,  ${\tau}^{TP}(L |0)$,  (circles) and  for
 transition paths starting from the right, ${\tau}^{TP}(0|L)$, (triangles). 
        The horizontal dashed line depicts the  force-free transition path time ${\tau}^{TP}( L | 0) D/L^2 =1/6$.
}
\end{figure*}


The transition path time reads
\begin{eqnarray}
{\tau}^{TP} \left( L |0 \right)
&=& {L^2 \int_{0}^{\sqrt{U}} \text{d} y ~ y^2  e^{-y^2} F_{2,2}(y^2)  \over D \sqrt{\pi} U \text{erf}(\sqrt{U})}.
\label{eq:tp2_half_harm}
\end{eqnarray}
For small $U$  we find the asymptotic  expression
\begin{eqnarray}
{\tau}^{TP} \left( L |0 \right) D/L^2
&\approx& {1 \over 6} - {U \over 90} - {2 U^2 \over 945},
 \label{eq:tp_half_harm_ap1}
\end{eqnarray}
while  for large $U$ we find
\begin{eqnarray}
{\tau}^{TP} \left( L |0 \right) D/L^2
&\approx& {\ln U  \over 4 U }.\label{eq:tp_half_harm_ap2}
\end{eqnarray}

Figure \ref{fig:ham_tpt_half} depicts ${\tau}^{TP}(L |0)$ as function of the barrier height $U$.
In Fig. \ref{fig:ham_tpt_half}-(a) we show, on  double logarithmic scales, 
 the numerically integrated ${\tau}^{TP}(L |0)$ from Eq. \eqref{eq:tp2_half_harm} by the solid red curve and compare with the asymptotic 
  expressions  Eqs. \eqref{eq:tp_half_harm_ap1} and \eqref{eq:tp_half_harm_ap2} (dashed curves).
         In Fig. \ref{fig:ham_tpt_half}-(b) we show ${\tau}^{TP}(L |0)$ from Eq. \eqref{eq:tp2_half_harm}   on a log-linear scale, 
         the symbols show BD simulation results.
        The solid blue curves in Fig. \ref{fig:ham_tpt_half} depict the Kramers' mean first-passage time, which is given by 
\begin{eqnarray}
{\tau}^{KFP} \left( L |0 \right)
&=&
{L^2 \over D}\left[ {{\pi  \text{erf}\left(\sqrt{U}\right) \text{erfi}\left(\sqrt{U}\right)} \over 4 U} - { F_{2,2}(-U) \over 2} \right],\label{eq:kfp_half_harm}\nonumber\\
\end{eqnarray}
and has the leading order expression
\begin{equation} \label{eq:kfp_half_harm_app1}
{\tau}^{KFP} \left( L |0 \right) = \frac{\sqrt{\pi} L^2}{4D}\frac{e^{U}}{U^{3/2}},
\end{equation}
 for large $U$.

The transition path time ${\tau}^{TP}(L |0) D/ L^2$ is nonmonotonic and is maximal for finite $U$  around $U \approx -21/8$,
 implying that transition paths that move down a weak harmonic ramp are slower  than in the force-free case.
For large $| U |$, ${\tau}^{TP}( L |0)$ decreases, similar  to the   linear potential case  shown in Fig. \ref{fig:4}-(a).
In contrast, the Kramers' mean first-passage time ${\tau}^{KFP}( L  | 0)$ exponentially increases as $U$ increases.


%
For the transition path shapes we find
\begin{widetext}
\begin{eqnarray}
{\tau}_{shape}^{TP}\left( {x}_0 | 0 \right)
&=&
{\tau}^{TP}\left( L | 0 \right) 
 -
{ \sqrt{\pi} L^2 \over 2 D U} \int_{0}^{y_0} \text{d}y e^{y^2} \text{erf}(y) \left[ 1+ \frac{\text{erf}(y)}{\text{erf}(y_0)} -2 \frac{\text{erf}(y)}{\text{erf}(\sqrt{U})}\right],\label{eq:harm_half_shape2}\\
{\tau}_{shape}^{TP}\left( {x}_0 | L \right)
&=&
{\tau}^{TP}\left( L | 0 \right) 
 -
{ \sqrt{\pi} L^2 \over 2 D U} \int_{y_0}^{\sqrt{U}} \text{d}y e^{y^2} 
\frac{ \left[ \text{erf}(\sqrt{U}) - \text{erf}(y) \right] \left[  \text{erf}(y_0) - \text{erf}(y) \right] }{ \text{erf} (y_0) - \text{erf} (\sqrt{U})}
,\label{eq:harm_half_shape3}
\end{eqnarray}
\end{widetext}
where $y_0 \equiv \sqrt{U}(1 - {x_0}/{L})$.

\begin{figure}
\centering
\includegraphics[width=0.49\textwidth]{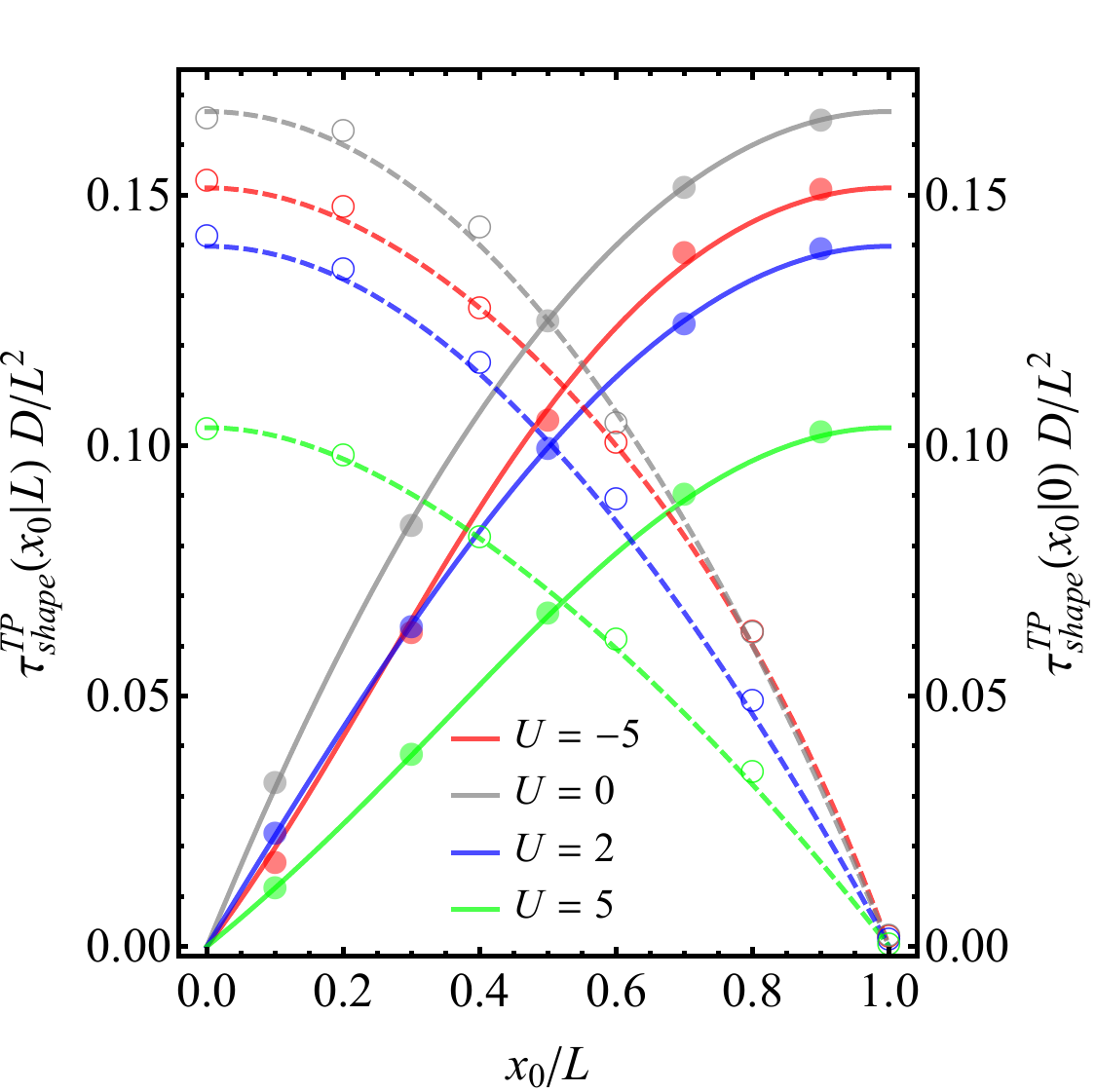}
\caption{\label{fig:harm_half_tpt_shape} Mean transition path shapes starting from the left, 
${\tau}_{shape}^{TP}({x}_0 | 0)$ (solid curves) from  Eq. \eqref{eq:harm_half_shape2}, 
and mean transition path shapes starting from the right,  ${\tau}_{shape}^{TP}({x}_0 | L)$ (broken curves) from Eq. \eqref{eq:harm_half_shape3}, 
for different values of the barrier height $U$ of the harmonic ramp ${F}({x}) = U {x}(2-{x/L})/L$.
 The symbols show the corresponding BD simulation results.
}
\end{figure}
 Figure \ref{fig:harm_half_tpt_shape} depicts the transition path shapes starting from the left,
 ${\tau}_{shape}^{TP}( {x}_0 | 0)$ (solid curves) from Eq. \eqref{eq:harm_half_shape2}, and starting from the right, 
 ${\tau}_{shape}^{TP}( {x}_0 | L)$ (broken curves) from Eq. \eqref{eq:harm_half_shape3}, for different values of the barrier height $U$.
        The symbols show the corresponding results from  BD simulations.
Note that the transition path shapes ${\tau}_{shape}^{TP}( {x}_0 | 0)$ and ${\tau}_{shape}^{TP}( {x}_0 | L)$ at constant 
$U$ are asymmetric with respect to the exchange of starting and end positions, due to the asymmetry of the barrier potential
(this becomes clear by comparing the mean shapes for $U=0$ (grey line) and for $U=-5$ (red line) starting from the left boundary
and starting from the right boundary).

\section{Conclusion}

Based on the one-dimensional Fokker-Planck equation, 
we develop the theoretical formalism to calculate mean shapes of  transition paths and of Kramers' first-passage paths
for arbitrary free energy and diffusivity landscapes. 
We use a combination of  the backward and forward Fokker Planck approaches  to derive explicit expressions for transition and first-passage path shapes.
To clarify the interpretation of our results, we also present  convolution expressions for  the distribution functions of transition path and passage times. 
We show that the mean shape of Kramers' first-passage paths is identical to the shape of transition paths  shifted by a constant.
Based on our analytic  theory, we present mean shapes  for several simple model potentials. We illustrate our results by trajectories generated from 
 Brownian dynamics simulations.
 Interestingly, transition path shapes are intrinsically asymmetric, they start out with finite velocity and reach the target position with infinite velocity,
 which is easily understood from our sum rules for transition path and passage  times.

The transition path shapes we predict can be compared straightforwardly with simulations for proteins that undergo folding and unfolding events
and will allow for a crucial test of the assumptions underlying the projection onto a one-dimensional reaction coordinate. 
 With  further developments of experimental single-molecule techniques, our results for the  transition path shapes can also be compared with
 experimental results in the future.
 For such a comparison, note that a reflecting boundary condition at $x=x_A$, as used in our calculations, is typically not present
 in molecular dynamics simulations nor in experiments. To apply our formulas, one can easily shift the reflecting boundary conditions to a position
 where the trajectory never visits. Alternatively, one can cut out all trajectory sections that visit the region behind the reflecting boundary condition
 and merge the remaining trajectory parts with  a continuous concatenated time, which is valid in the limit of vanishing memory and effective mass.
 


\section*{Acknowledgements}
The authors thank Bill Eaton for stimulating discussions.
Financial support from the DFG (SFB 1078)  is acknowledged.



\end{document}